\documentclass[11pt,preprint]{aastex}

\newcommand{\eg}{{\it e.g.}}    % define 'for example'
\newcommand{\ie}{{\it i.e.}}    % define 'that is'
\newcommand{\etal}{et al.}      % define 'and others'
\newcommand{\inv}{$^{-1}$}
\newcommand{\kms}{km~s\inv}

\newcommand{\gapsize}{\Delta\vpec}
\newcommand{\maxgapsize}{\Delta v_{\rm max}}
\newcommand{\Rvirial}{R_{200}}
\newcommand{\vpec}{v_{\rm pec}}
\newcommand{\vpeci}{v_{{\rm pec},i}}

\newcommand{\Cl}{Cl\,}          % leave space between 'Cl' and number
\newcommand{\MS}{MS\,}          % leave space between 'Cl' and number
\newcommand{\veldisp}{\sigma_p} % projected velocity dispersion
\newcommand{\woi}{\Cl0016+16}
\newcommand{\wov}{\MS0451--03}
\newcommand{\woviii}{{\Cl}J1324+3011}
\newcommand{\wovii}{\MS1054--03}
\newcommand{\wx}{{\Cl}J1604+4304}
\slugcomment{Revised draft, 2014-Mar-03}

\shorttitle{Transitional Populations in Galaxy Clusters}
\shortauthors{Crawford \etal}

\begin{document}

\title{Spatial and Kinematic Distributions of Transition Populations
  in Intermediate Redshift Galaxy Clusters\altaffilmark{1}}

\author{Steven M. Crawford}
\affil{SAAO, P. O. Box 9, Observatory 7935, Cape Town, South Africa}
\email{crawford@saao.ac.za}

\author{Gregory D. Wirth}
\affil{W. M. Keck Observatory, 65-1120 Mamalahoa Hwy, Kamuela, HI 96743}
\email{wirth@keck.hawaii.edu}

\and

\author{Matthew A. Bershady} 
\affil{Department of Astronomy, University of Wisconsin, 475 North
  Charter Street, Madison, WI 53706}
\email{mab@astro.wisc.edu}

\altaffiltext{1}{ Based in part on data obtained at the W.~M. Keck
  Observatory, which is operated as a scientific partnership among the
  California Institute of Technology, the University of California and
  the National Aeronautics and Space Administration. The Observatory
  was made possible by the generous financial support of the W.~M.
  Keck Foundation.}

\begin{abstract}
  
  We analyze the spatial and velocity distributions of confirmed
  members in five massive clusters of galaxies at intermediate
  redshift ($0.5 < z < 0.9$) to investigate the physical processes
  driving galaxy evolution.  Based on spectral classifications derived
  from broad- and narrow-band photometry, we define four distinct
  galaxy populations representing different evolutionary stages: red
  sequence (RS) galaxies, blue cloud (BC) galaxies, green valley (GV)
  galaxies, and luminous compact blue galaxies (LCBGs).  For each
  galaxy class, we derive the projected spatial and velocity
  distribution and characterize the degree of subclustering.  We find
  that RS, BC, and GV galaxies in these clusters have similar velocity
  distributions, but that BC and GV galaxies tend to avoid the core of
  the two $z\approx0.55$ clusters.  GV galaxies exhibit subclustering
  properties similar to RS galaxies, but their radial velocity
  distribution is significantly platykurtic compared to the RS
  galaxies.  The absence of GV galaxies in the cluster cores may
  explain their somewhat prolonged star-formation history.  The LCBGs
  appear to have recently fallen into the cluster based on their
  larger velocity dispersion, absence from the cores of the clusters,
  and different radial velocity distribution than the RS galaxies.
  Both LCBG and BC galaxies show a high degree of subclustering on the
  smallest scales, leading us to conclude that star formation is
  likely triggered by galaxy-galaxy interactions during infall into
  the cluster.

\end{abstract}

\keywords{Galaxies: distances and redshifts -- galaxies: photometry --
galaxies: galaxy clusters} 

\section{Introduction}
\label{sxn-intro}

A substantial body of astronomical research conducted over the last
several decades has established that environment strongly modulates
galaxy properties such as star formation rate \citep{gomez03} and
gives rise to the so-called morphology-density relationship
\citep{dressler80} seen in the nearby Universe.  These studies paint a
picture in which dense cluster environments transforms blue,
star-forming galaxies into red, queiscent galaxies at much faster
rates than in the low-density field. However, the responsible
mechanisms and timescales for these transformations in clusters remain
a matter of debate \citep{bg06}.

An important factor affecting the evolution of individual cluster
galaxies is thought to be the nature of their orbits within the
cluster: highly-radial orbits are far more susceptible to effects such
as ram-pressure stripping that quench star formation \citep{solanes01,
  iannuzzi12}.  Certain studies of low-redshift clusters find that
star-forming galaxies are more likely than passive galaxies to occupy
radial orbits \citep{biviano97, mahdavi99, biviano04}, suggesting that
recent infall accounts for today's star-forming cluster galaxies.
Other studies report contradictory results for nearby clusters
\citep{ramirez98, hwang08} and possible evolution with redshift for
the quiescent population \citep{ramirez00, biviano09}.

Motivated by a desire to clarify this picture, we seek to quantify how
location and kinematics affect the evolution of cluster galaxies.  To
do so, we examine the available phase-space properties (\ie, the
projected spatial and velocity distributions) of different
optically-identified populations in galaxy clusters at intermediate
redshift.  By identifying galaxies undergoing rapid evolution and
studying their phase-space properties, we hope to gain insight into
how galaxies in clusters evolve from the actively star-forming phase
into passively evolving systems.  Well-known examples of these
transitional galaxy types include post-starburst (``E+A'') galaxies
\citep{dg83} and ultra-luminous infrared galaxies
\citep[ULIRGs,][]{soifer84}.  

The present study focuses on two other evolving galaxy populations
that we can readily identify on the basis of their photometric
properties alone: ``green valley'' galaxies and luminous compact blue
galaxies.  Green valley galaxies \citep[GV,][]{wyder07} occupy a
region in optical color-magnitude space straddling the dividing line
between the well-defined red sequence and the blue cloud
\citep{wyder07}.  Luminous compact blue galaxies (LCBG) are an extreme
class of star-forming galaxies first identified in optical multi-color
QSO surveys \citep{koo88}.  The LCBG class is characterized by small
physical sizes \cite[$R_e \sim 2$~kpc based on HST imaging;][]{koo94} and
narrow line widths 
\citep[$30<\sigma<80$~\kms\ based on Keck spectroscopy;][]{guz97}.

In all such transitional populations, major changes in their
properties --- including morphology, luminosity, and color --- are
believed to occur quickly relative to the age of the Universe;
consequently, galaxies transiting between active and quiescent phases
are relatively rare.  Their remarkably low space densities in the
field make it difficult to gather samples sufficiently large to enable
meaningful analysis; however, the higher space density of galaxies in
clusters permits the study of statistically-significant sample sizes
providing snapshots of galaxy evolution at specific cosmic epochs.

The GV galaxies are believed to represent a transitory population
between the GC and RS phases \citep{coil08, salim09}.  As with the
colors of the galaxies, the morphology (in terms of concentration and
asymmetry) of GVs are intermediate between these two classes
\citep{mendez11}.  The GV class has not been well studied to date as a
distinct cluster population, although spectroscopically-identified
post-startburst galaxies (\eg, ``E+A'' galaxies) have received much
attention in the literature \citep[\eg,][]{barger96,tran03, muzzin12}.

Our interest in LCBGs stems from the suggestion that in the field
these galaxies rapidly evolve from intermediate redshifts to form
dwarf ellipticals in present day clusters. In this scenario, given the
relative abundance of dwarf ellipticals in clusters, a surfeit of
LCBGs might be expected in intermediate redshift clusters.
\cite{cra06,cra11} found that the number density of LCBGs was enhanced
in cluster environments by a factor of $749\pm116$ at $z\sim0.5$, but the cause of 
this increased density of relatively rare objects is not understood.

In a series of papers \citep{cra06,cra09, cra11} we have explored the
red (quiescent) and blue (star-forming) populations in a half-dozen
rich clusters over the redshift range $0.5<z<1$. Our aim has been to
understand how these general populations evolve differentially with
respect to the field. In this work, we compare and contrast the
phase-space distribution of various galaxy populations to obtain
insight into the morphology-density relationship, star-formation
quenching, and population transformation in clusters. We isolate the
LCBG and GV transitional classes from the global red and blue cluster
populations based on their colors and radial profiles.  Our work
builds upon a deep optical imaging survey from the WIYN 3.5~m
telescope\footnote{The WIYN Observatory is a joint facility of the
  University of Wisconsin-Madison, Indiana University, Yale
  University, and the National Optical Astronomy Observatories.}  and
spectroscopic follow-up from the DEIMOS spectrograph \citep{fab03} on
the Keck~II Telescope of the W.~M. Keck Observatory.  The parameters
of each of our clusters is listed in Table~\ref{tab_cluster_params}.
Details of the observations appear in \cite{cra09} and \cite{cra11}.

\begin{deluxetable}{lcllrrrr}
\tabletypesize{\footnotesize}
  \tablewidth{0pc}
  \tablecaption{Summary of Fields\label{tab_cluster_params}}
  \tablehead{
    \colhead{Field}         &
    \colhead{WLTV ID}         &
    \colhead{$\alpha$} &
    \colhead{$\delta$} &
    \colhead{$z$} &
    \colhead{$\veldisp$} &
    \colhead{$\Rvirial$} &
    \colhead{$\Rvirial$} \\
    &
    &
    \colhead{(J2000)} &
    \colhead{(J2000)} &
    &
    \colhead{(\kms)} &
    \colhead{(Mpc)} &
    \colhead{($\arcsec$)} \\
    \colhead{(1)} &
    \colhead{(2)} &
    \colhead{(3)} &
    \colhead{(4)} &
    \colhead{(5)} &
    \colhead{(6)} &
    \colhead{(7)} &
    \colhead{(8)} 
  }
  \startdata
  MS~0451-03 & 
w05 & 
04:54:10.8 & 
$-$03:00:51 & 
 $     0.5389 \pm     0.0005 $  & 
 $         1328 \pm           47 $  & 
 $       2.45 \pm       0.09 $  & 
 $          386 \pm           13 $  \\
Cl~0016+16 & 
w01 & 
00:18:33.6 & 
$+$16:26:16 & 
 $     0.5467 \pm     0.0006 $  & 
 $         1490 \pm           80 $  & 
 $       2.74 \pm       0.15 $  & 
 $          428 \pm           23 $  \\
Cl~J1324+3011 & 
w08 & 
13:24:48.8 & 
$+$30:11:39 & 
 $     0.7549 \pm     0.0007 $  & 
 $          806 \pm           85 $  & 
 $       1.31 \pm       0.14 $  & 
 $          178 \pm           18 $  \\
MS~1054-03 & 
w07 & 
10:56:60.0 & 
$-$03:37:36 & 
 $     0.8307 \pm     0.0006 $  & 
 $         1105 \pm           61 $  & 
 $       1.72 \pm       0.10 $  & 
 $          225 \pm           12 $  \\
Cl~J1604+4304 & 
w10 & 
16:04:24.0 & 
$+$43:04:39 & 
 $     0.9005 \pm     0.0014 $  & 
 $         1106 \pm          167 $  & 
 $       1.65 \pm       0.25 $  & 
 $          211 \pm           32 $  \\

  \enddata
  \tablecomments{(1)~Standard cluster field name; (2)~internal
  designation for each of the clusters; (3) and (4)~celestial
  coordinates of the adopted cluster center defined by Brightest
  Cluster Galaxy; (5)~measured cluster redshift; (6)~measured cluster
  projected velocity dispersion based on the biweight estimator of scale; 
  (7)~cluster virial radius derived from
  $\veldisp$  for our
  adopted cosmology; (8)~cluster virial radius in angular units.}
%  Notes.--(1)~Cluster also known by the identifier MS~0016.9+1609. 
%  (2)~Cluster centroid position from \Chandra\ ACIS X-ray analysis
%  of \cite{mau08}. (3)~Cluster centroid position from \emph{ROSAT} HRI
%  X-ray analysis of \cite{neu00}.  (4)~Cluster centroid position from
%  ?? (ask Steve where his numbers on th project page came from).  (5)~Cluster centroid position from
%  \Chandra\ ACIS analysis of \cite{koc09}.
\end{deluxetable}

Our paper is organized as follows: In \S2, we describe our photometric
classification scheme for different populations; in \S3 we describe
our method for culling a reliable cluster sample for objects with
spectroscopic redshifts; in \S4, we analyze different projections of
the accessible phase-space distribution; in \S5, we consider the
subclustering of the different populations; we discuss our results in
\S6 and summarize our findings in \S7. Throughout this paper, we employ
the following acronyms to refer to different classes of galaxies
defined in \S2: red sequence (RS), blue cloud (BC), green valley (GV),
and luminous compact blue galaxies (LCBGs).

\section{Photometric Classification}
\label{sxn-photclass}

We separate the galaxies in each field into distinct samples
regardless of cluster membership based purely on their {\it
  rest-frame} photometric properties: color, luminosity, and surface
brightness.  Building on the work of \cite{cra11}, we define four
galaxy samples:

\begin{itemize}

\item Red sequence (RS) galaxies are defined as objects redder than
  dividing line between blue and red objects, which we define as $U-B
  = -0.032 \times (M_B + 21.52) + 0.204 $ following
  \citet{wilmer06}.\footnote{This definition is based on a $-0.25$ mag
    shift in the zeropoint of the color-magnitude relationship at
    intermediate redshifts.}

\item Blue cloud (BC) galaxies are taken to be bluer than this same
  dividing line.

\item Green valley (GV) galaxies occupy an 0.2-mag-wide strip in color
  centered on the line between RS and BC galaxies cited above; the
  adopted definition is comparable to others found in the literature
   \citep[\eg,][]{coil08}.
   
 \item Luminous compact blue galaxies (LCBGs) are a subset of the BC
   class, defined as having $B-V<0.5$, $\mu_B < 21~\mbox{mag
     arcsec}^{-2}$, and $M_B < -18.5$ \citep{garland04, cra06}.
\end{itemize}

\begin{figure}
  \includegraphics[angle=270, scale=0.75]{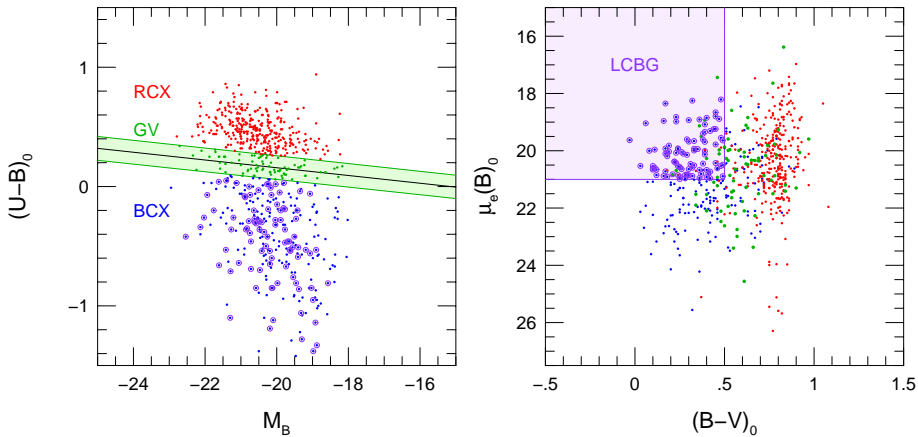}
  \caption{\label{fig_cmd_ex} Illustration of photometric selection
    criteria for RSX, GV, BCX, and LCBG classifications in $(U-B)_0$
    versus $M_B$ (left) and the mean $B$-band surface-brightness
    within the half-light radius ($\mu_e(B)_0$) versus $(B-V)_0$.  All
    quantities are expressed in the cluster rest frame.  Panels show
    all bona fide cluster members with spectroscopic redshifts in our
    sample.  }
\end{figure}

A shortcoming of the above definitions is that the GV class overlaps
the BC and RS classes in color-magnitude space, making the
classification of GV galaxies ambiguous.  In order to enable
statistical comparisons between independent samples of galaxies, we
choose to redefine the RS and BC classes to exclude the GV galaxies;
thus, in our ensuing analysis we define ``exclusive-RS'' (RCX) and
``exclusive-BC'' (BCX) subsamples to be the RS and BC subsamples
excluding the intermediate-color GV sample.  Table~\ref{tab_classes}
summarizes our adopted galaxy types.  An example of the selection for
different sources is shown in Figure~\ref{fig_cmd_ex}.

%----------------------------------------
% tab_classes
%----------------------------------------
\begin{deluxetable}{llll}
\tabletypesize{\footnotesize}
  \tablewidth{0pc}
  \tablecaption{Adopted Galaxy Classes
    \label{tab_classes}
  }
  \tablehead{
    \colhead{Sample} &
    \colhead{Color} &
    \colhead{Surface Brightness}  &
    \colhead{Luminosity}
  }
  \startdata
  RSX & $U-B > -0.032 \times (M_B + 21.52) + 0.304$ & $\cdots$ & $\cdots$ \\
  GV & $U-B = -0.032 \times (M_B + 21.52) + 0.204 \pm 0.1$ & $\cdots$ & $\cdots$ \\
  BCX & $U-B < -0.032 \times (M_B + 21.52) + 0.104$ & $\cdots$ & $\cdots$ \\
  LCBG &   $B-V<0.5$ & $\mu_B < 21~\mbox{mag arcsec}^{-2}$ & $M_B < -18.5$\\
  \enddata
\end{deluxetable}

%% Section 2
\section{Cluster Membership}
\label{sxn-member}

Since our ultimate goal is to compare and contrast the properties of
various galaxy types within clusters, we must define which galaxies
are bona fide members of their respective clusters. To avoid biasing
the samples either for or against any type of galaxy, we require a
method which is blind to the intrinsic properties of the galaxy (\eg,
color, morphology, or spectral type).  We adopt the customary strategy
of applying a test called the ``shifting gapper,'' which analyzes the
velocity distribution of galaxies in each field to determine which
targets belong to the cluster \citep[\eg,][]{fad96,lop09}.  We also
apply a radial cutoff to exclude galaxies lying more than a certain
distance from the cluster center in projected space.  Due to the
interdependence in the determinations of the velocity dispersion and
the radial cutoff, we developed an iterative approach to solve the
problem, as described next.

\label{sxn-shifting-gapper}

\subsection{Velocity Cut Algorithm} 

To determine which galaxies were members of a given cluster, we began
by employing the ``shifting gapper'' method to define the cutoff in
radial velocity, but modified the algorithm in two important ways:
\begin{itemize}
  
\item The velocity gap size used to distinguish members from
  interlopers is based on considering the $N$ nearest neighbors in
  clustocentric radius, rather than by constructing bins of size
  0.4~Mpc in clustocentric distance or having at least $N$ members.
  Our approach eliminates the discontinuities inherent in a
  binning-based analysis and ensures that the same number of galaxies
  is used in each computation of the gap size.
  
\item We employed the absolute value of the peculiar velocity in the
  analysis, rather than the raw value, since clusters are
  approximately symmetrical in radial velocity space.

\end{itemize}

\begin{figure}[p]
  \epsscale{0.7}
  \plotone{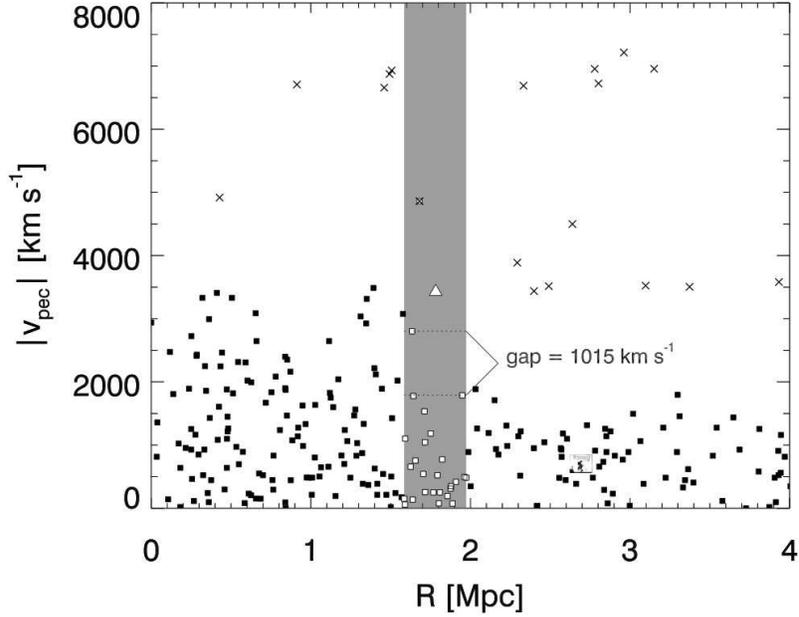}
  \caption{
    Application of the ``shifting gapper'' algorithm for determining
    cluster membership.  The distribution of absolute peculiar
    velocity $|\vpec|$ as a function of clustocentric distance $R$ is
    shown for galaxies in one of our fields.  To determine whether a
    particular galaxy (indicated by the open triangle) at ($R_i,
    \vpec,i$) is a member of the cluster, we form a ``neighbor''
    sample consisting of the 30 galaxies with $R$ values closest to
    $R_i$; such galaxies occupy the shaded region centered on the
    galaxy of interest are are indicated by open squares.  Within this
    subsample, we search for the largest gap in the distribution of
    $|\vpec|$, considering only $|\vpec| < |\vpec,i|$.  In the case of
    this particular target, we find the maximum gapsize to be
    1015~\kms.  Since this gap size exceeds the maximum permitted
    value for cluster members (taken to be 1000~\kms), this galaxy
    fails the membership criterion and is rejected as a cluster
    member.  Other objects which were similarly rejected are shown as
    X symbols, while those classified as members are shown as filled
    or open squares.  }
  \label{fig_shifting_gapper_demo}    
\end{figure}

Given the distribution of galaxy clustocentric radii, $R_i$, and the
corresponding peculiar velocities, $\vpeci$, any given galaxy is
classified as either a cluster member or non-member as follows:
\begin{enumerate}
  
\item Locate the $N$ galaxies whose radii $R$ from the cluster center
  are closest to $R_i$.  We take $N$ to be 30.
  
\item Within this sample of $N$ targets, identify the subset of those
  targets which have $|\vpec| < |\vpeci|$ and measure $\gapsize$,
  the biggest gap in the distribution of absolute peculiar velocity
  within this sample.
  
\item If the gap size $\gapsize$ is less than a selected maximum value
  $\maxgapsize$, then this object is accepted as a member; otherwise,
  it is classified as a non-member.

\end{enumerate}
The procedure is depicted graphically in
Figure~\ref{fig_shifting_gapper_demo}.

\subsection{Radial Cut Algorithm}

In addition to the velocity cut, it is necessary to define an outer
limit in projected radius for cluster members.  To exclude galaxies
which are not bound to the cluster, we required targets to lie within
the virial radius of the cluster, estimated via
\begin{equation}
  \Rvirial \equiv\sqrt{3} \frac{\veldisp}{10H(z)},
  \label{eq:r200}
\end{equation}
where $\Rvirial$ is the radius at which the density of the cluster is
$200\times$ the critical density of the Universe.  This follows directly
for a singular isothermal sphere ($\rho \propto R^{-2}$), where
$\veldisp$ is the projected velocity dispersion of the cluster and
$H(z)$ is the value of the Hubble parameter at the cluster
epoch\footnote{We adopt $H_0=70 {\rm km s}^{-1}{\rm Mpc}^{-1}$,
  $\Omega_m=0.3$, and $\Omega_\Lambda=0.7$ throughout.}.

\subsection{Iterative Estimation of Cluster Parameters}
The velocity and radial criteria described above are interdependent.
The value of $\veldisp$ (from which $\Rvirial$ is derived) depends on
which galaxies are included in the sample; thus, to measure $\veldisp$
we should exclude those (presumably unbound) targets lying outside
$\Rvirial$.  However, the formula for $\Rvirial$ requires a measured
value of $\veldisp$; hence, the process is circular.  We thus adopted
an iterative approach to reject interlopers while simultaneously
deriving values for the cluster redshift $z$, terminal radius
$\Rvirial$, and projected velocity dispersion $\veldisp$.

Given a first estimate of the cluster redshift, which is derived by
finding a peak in the redshift distribution, the method proceeds as
follows:

\begin{enumerate}

\item We compute the peculiar velocities of all galaxies (measured in
  the cluster rest frame) relative to the nominal cluster redshift,
  then exclude all galaxies with peculiar velocities exceeding
  10,000~\kms\ in absolute value.

\item We employ our modified ``shifting gapper'' method
  (\S\ref{sxn-shifting-gapper}) to exclude interloping galaxies based
  on peculiar velocity.

\item Using the subset of galaxies which pass the interloper rejection
  test, we recompute the cluster redshift $z$ and projected velocity
  dispersion $\veldisp$ using robust, biweight-based estimates of
  location and scale \citep{bee90}.

\item Given the measured value of $\veldisp$, we compute $\Rvirial$ via
  eq.~\ref{eq:r200}, then convert this physical value to an angular
  measurement based on the cluster redshift and assumed cosmological
  parameters.

\item We repeat the previous 3 steps, now excluding targets which lie
  outside of $\Rvirial$ as well as those which fail the ``shifting
  gapper'' test.

\end{enumerate}

We continue this iterative procedure until the value of $\veldisp$ and
the number of cluster members stabilize within 1\% or until a maximum
25 iterations are completed.  In practice, convergence generally
occurs within five iterations.

\subsection{Parameter Values}
\label{param-values}

We now consider whether our results are sensitive to the particular
values we chose for the three key parameters in our process:
\begin{enumerate}

\item The initial peculiar velocity window size used to reject obvious
  outliers, taken to be $\pm10,000$~\kms\ (in the cluster rest frame).

\item The number of neighbors $N$ employed in the ``shifting gapper''
  algorithm to compute the gap size, taken to be 30.

\item The maximum permitted gap size employed in the ``shifting
  gapper,'' $\gapsize$, taken to be 1000~\kms.

\end{enumerate}

We explored this by using the \woi\ field dataset as a testbed to
investigate the dependency of the derived cluster redshift, cluster
velocity dispersion, and the number of cluster members on the values
of these parameters.

\begin{figure}
  \plotone{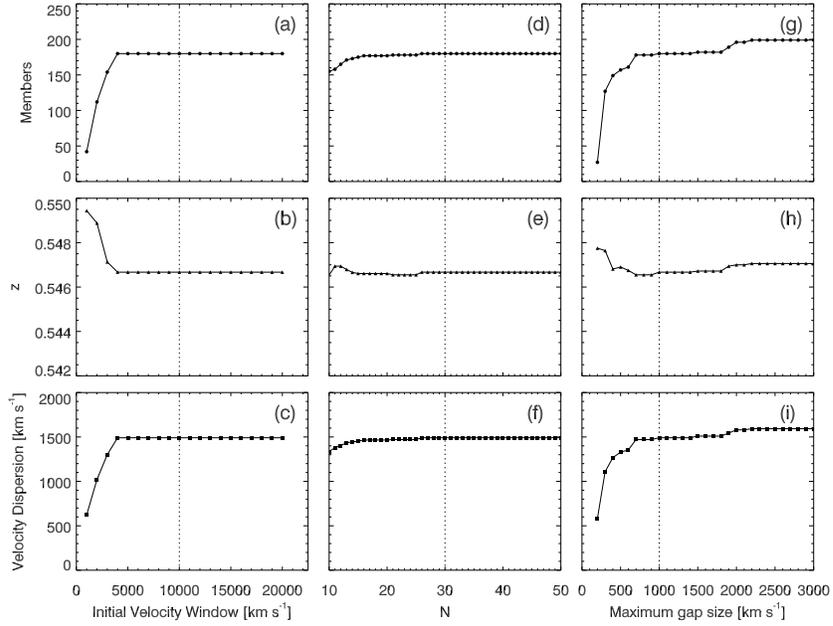}
  \caption{\label{fig_test_center_scale_iter_vary_all}
    Results of experiments we performed to determine optimal values of
    three key parameters in our cluster membership analysis.  Panels
    (a)--(c) show that varying the initial velocity window size
    had relatively little impact on the derived number of cluster
    members, central redshift $z$, and velocity dispersion; the vertical
    line indicates our adopted value of $\pm10,000$~\kms.
    Panels (d)--(f) illustrate that varying the number of neighbors
    considered when computing the velocity gap size had virtually no
    effect on the same derived cluster properties; vertical line marks
    our adopted value of $N=30$.  Panels (g)--(i) indicate that
    changing the maximum gap size parameter has little impact on these same
    derived properties; vertical line indicates our adopted value of
    $1000$~\kms. 
  }
\end{figure}

Figure~\ref{fig_test_center_scale_iter_vary_all} shows the results of
experiments in which we varied these parameters to gauge their
impact on our measurements of three key cluster properties: the
derived number of cluster members; the central redshift $z$ as
characterized by the biweight estimate of central location; the
derived line-of-sight velocity dispersion $\veldisp$, as estimated via
the biweight estimator of scale.  Panels (a)--(c) consider the effect
of changing the initial peculiar velocity window size.  Here, galaxies
with an absolute peculiar velocity greater than the specified value
were excluded from the subsequent calculation of the cluster
properties.  We repeated our analysis in the \woi\ field using values
of the window size ranging from 1000~\kms\ to 20,000~\kms.  Results
indicate that the number of derived cluster members, the cluster
redshift, and the cluster velocity dispersion are completely
insensitive to changes in the window size parameter over the range
4,000--19,000~\kms.  We conclude that 10,000~\kms\ is a reasonable
value for the window size.

Similarly, panels (d)--(f) show the effect on derived cluster
properties of varying the number of neighbor galaxies considered in
computing the gap size for our implementation of the ``shifting
gapper'' algorithm.  The plots indicate that values in the range
$26<N<50$ yield similar results for \woi.  We select
$N=30$ to preserve locality.

Finally, panels (g)--(i) illustrate the effect on derived cluster
parameters of varying the maximum permitted velocity gap size
($0<\gapsize<3000$~\kms) in the ``shifting gapper'' algorithm.  These
plots indicate that values in the range 800--1700~\kms\ yield similar
results.  We follow \cite{fad96} and select $\gapsize=1000$~\kms.

\section{Phase-space Distribution of Cluster Populations}
\label{sxn-phase}

For clues into what processes drive galaxy evolution in these
clusters, we investigated whether the star-forming and
non-star-forming populations have different spatial and/or velocity
distributions.  Position-velocity and position-position diagrams for
all galaxies with measured spectroscopic redshfts in each of the
cluster fields appear in Figures~\ref{fig_phase.w01} through
\ref{fig_phase.w10}. They identify which objects we consider
\emph{bona fide} cluster members based on the protocol defined in
\S\ref{sxn-member}. For these \emph{bona fide} members, the figures
also distinguish objects by photometric type within $\Rvirial$ or to
the limits of the areal coverage of our WIYN multi-band photometry
\citep{cra11}, whichever is smaller.  The figures show that the two
lower-redshift clusters in our sample are rich and well-defined,
whereas the higher-redshift clusters are less well sampled due to a
combination of being intrinsically poorer and having apparently
fainter galaxies \citep{lubin98, post01}.

\begin{figure}
  \epsscale{1.0}
  \plotone{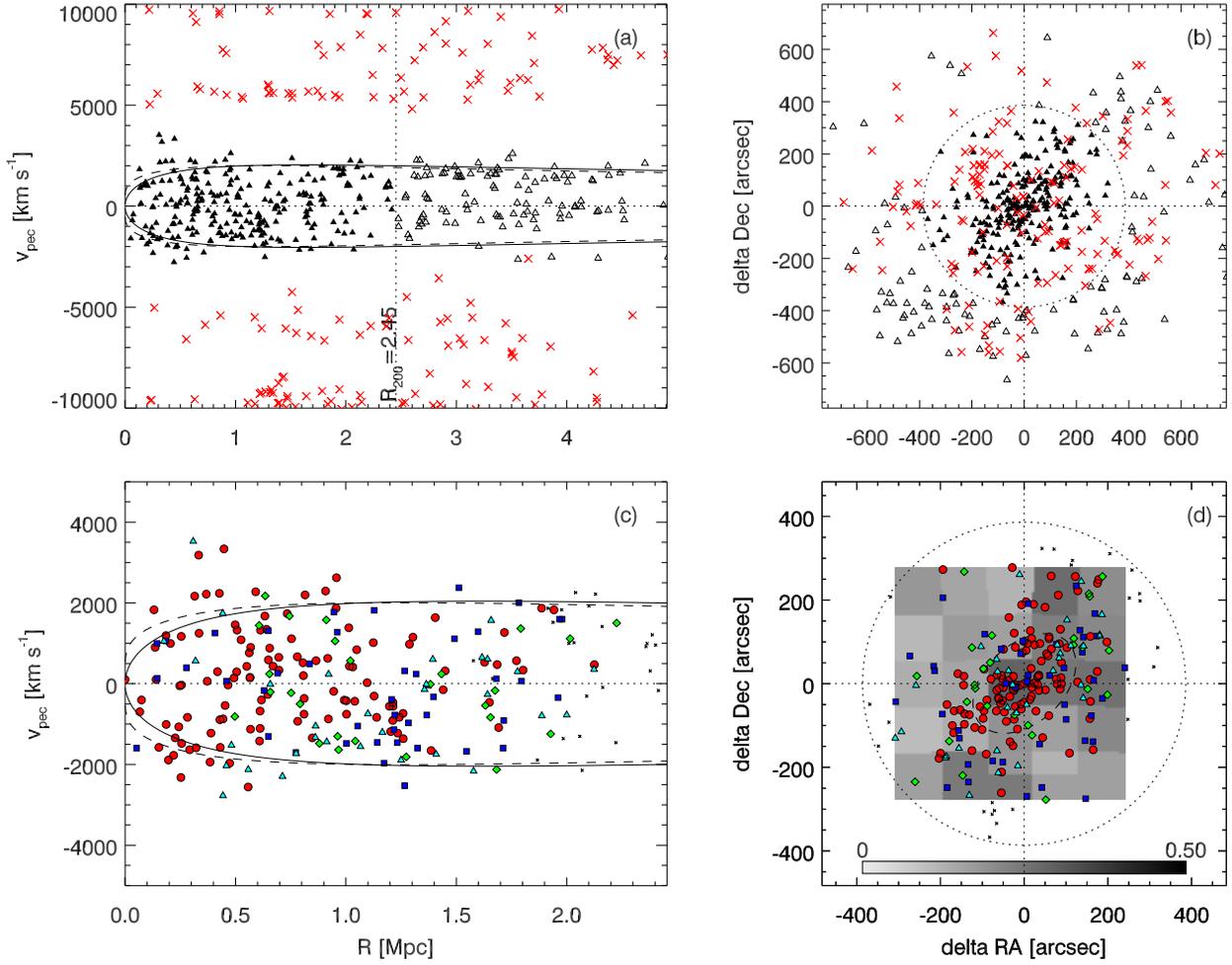}  
  \caption{Spatial and velocity distributions for targets in the
    \wov\ field.  (a)~Radius-velocity distribution for galaxies with
    respect to the cluster barycenter given in Table
    \ref{tab_cluster_params}.  Solid and dashed lines approximate the
    gravitational boundaries of the cluster based on the velocity
    dispersion in the literature, as described in the text.  Filled
    triangles indicate objects meeting membership criteria on velocity
    and radius. Open triangles denote targets meeting velocity
    criterion but falling outside of $\Rvirial$ (marked by dotted
    lines).  X symbols denote targets failing the ``shifting gapper''
    velocity cut. (b)~Projected sky distribution for targets in this
    field.  Symbols as in panel a.  (c)~Radius-velocity distribution
    for targets classified as cluster members out to $\Rvirial$.
    Color indicates spectral type (red=RSX, blue=BCX, green=GV,
    cyan=LCBG).  Black crosses indicate targets without predicted
    spectral types because they lie outside the field of view of the
    WIYN narrow-band photometry survey.  (d)~Projected sky
    distribution for cluster members.  Symbols as in panel (c). Grey
    background shading indcates relative completeness of the survey in
    that projected volume, with the color bar at bottom indicating the
    translation from color to completeness. The dotted circle
    represents $\Rvirial$ and the dashed ellipse indicates the
    completeness-weighted centroid and ellipticity of the galaxy
    distribution.}
\label{fig_phase.w05}
\end{figure}

\begin{figure}
  \plotone{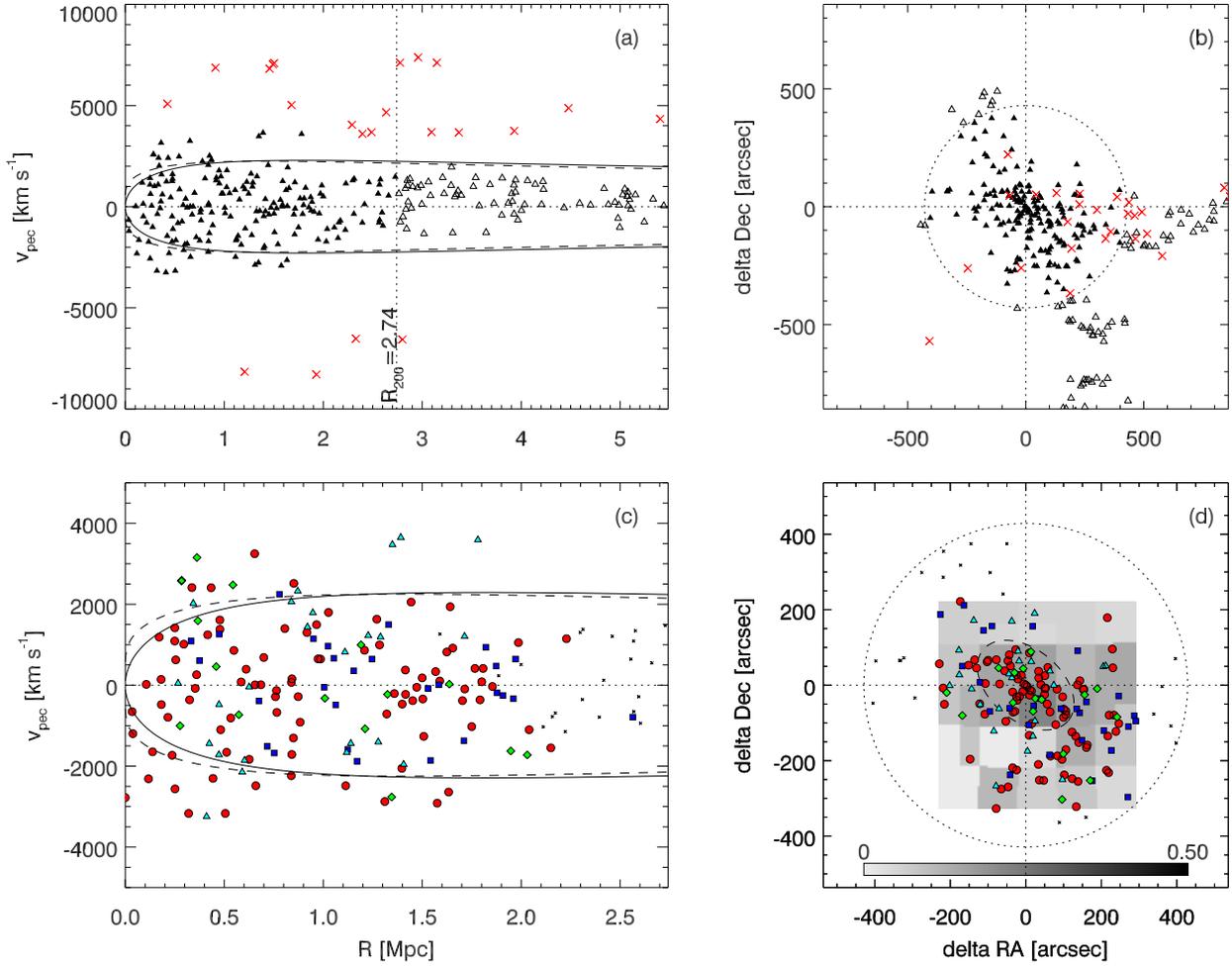}
  \caption{
    Radius-velocity and position-position distributions for our \woi\
    field, as described in Figure~\ref{fig_phase.w05}.}
\label{fig_phase.w01}
\end{figure}

\begin{figure}
  \plotone{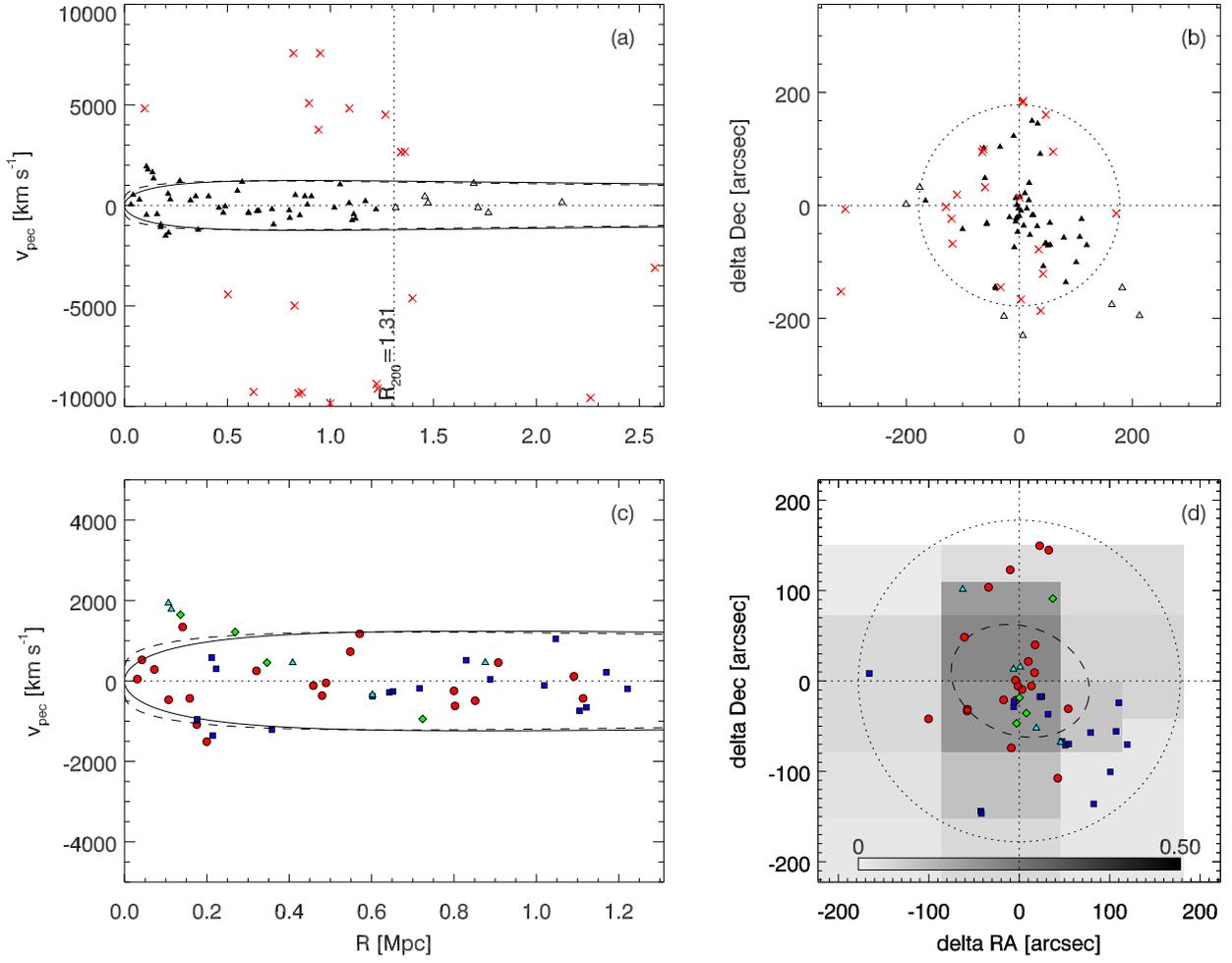}
  \caption{
    Radius-velocity and position-position distributions for our \woviii\
    field, as described in Figure~\ref{fig_phase.w05}.}
\label{fig_phase.w08}
\end{figure}

\begin{figure}
  \plotone{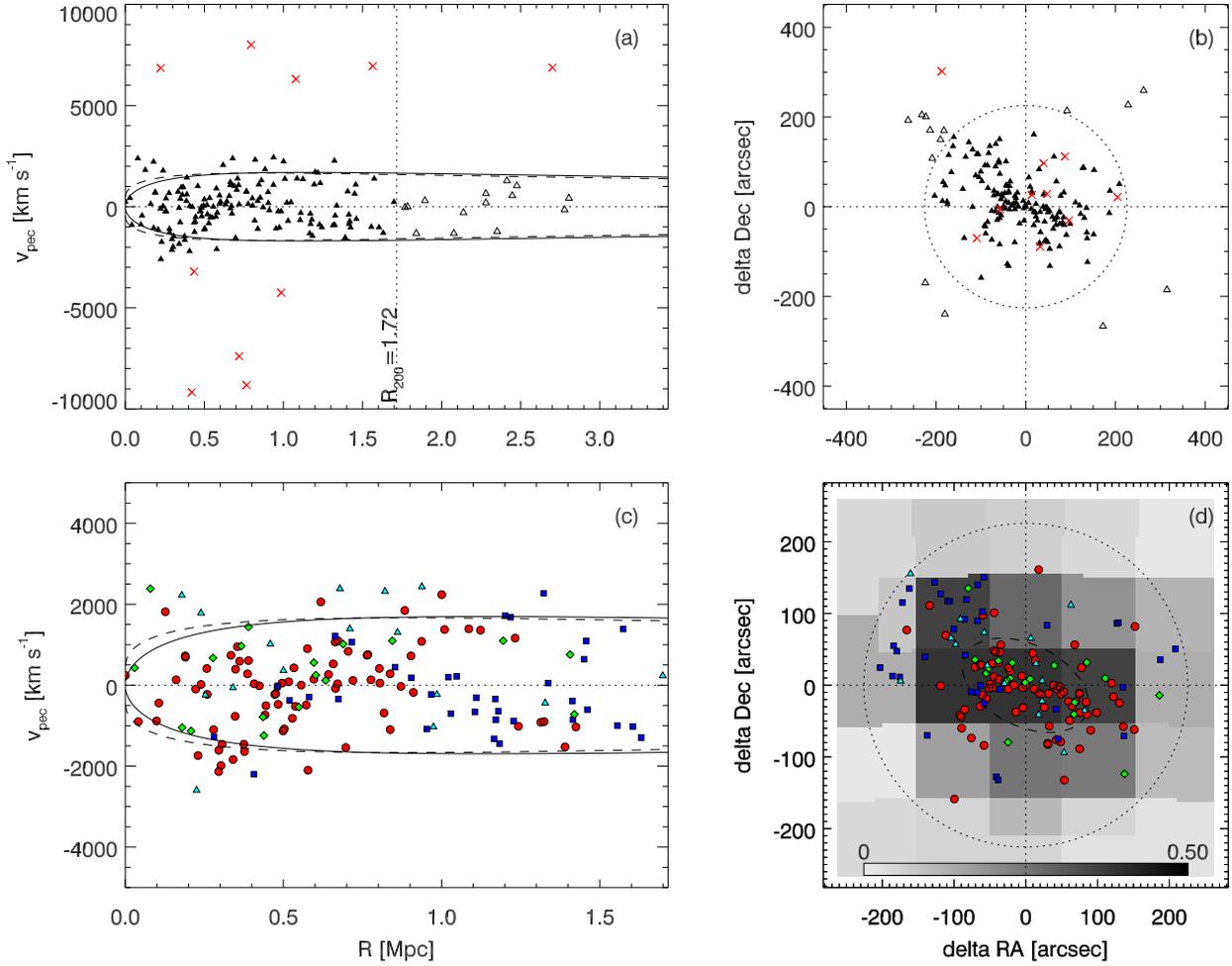}
  \caption{
    Radius-velocity and position-position distributions for our \wovii\
     field, as described in Figure~\ref{fig_phase.w05}.}
  \label{fig_phase.w07}
\end{figure}

\begin{figure}
  \plotone{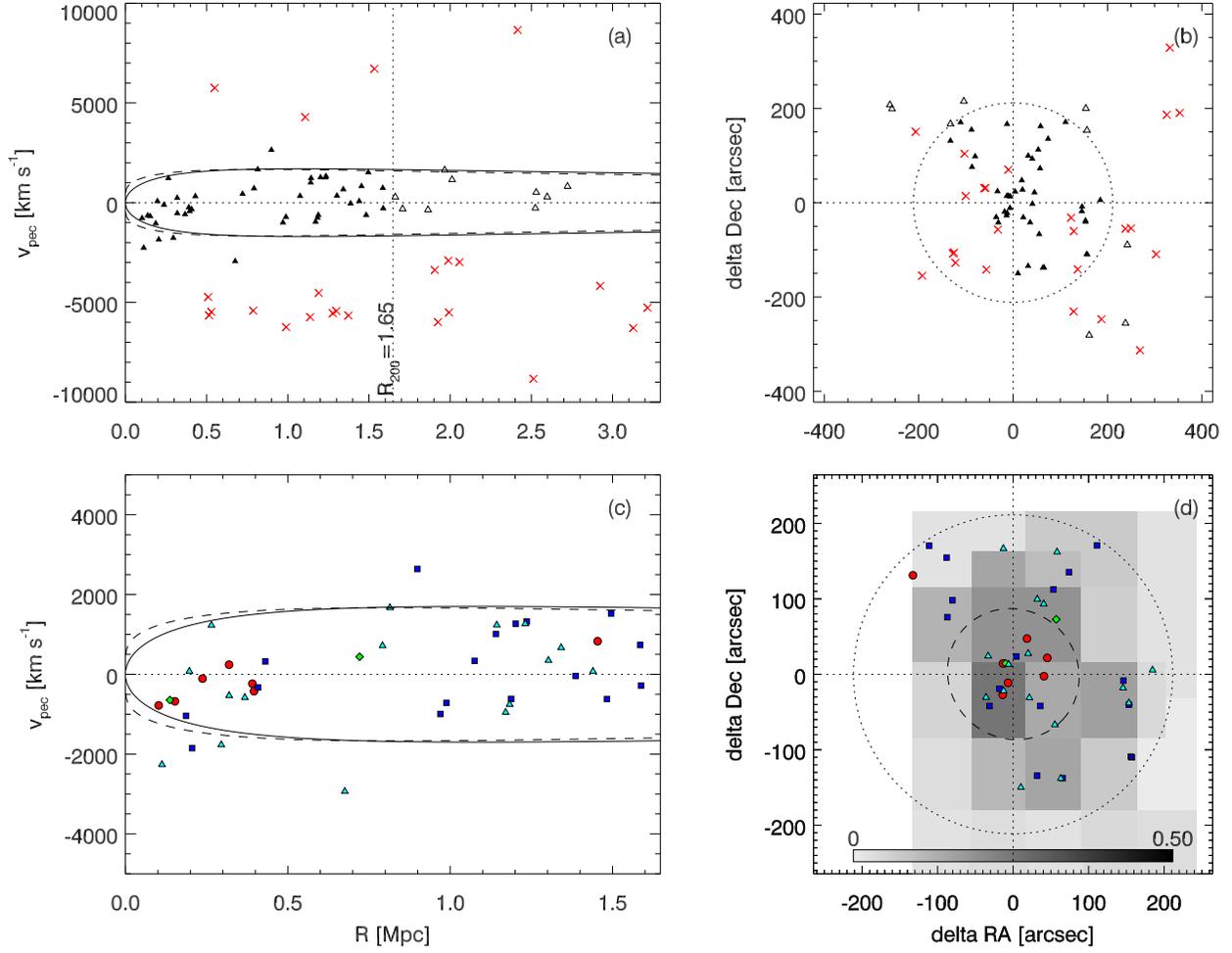}
  \caption{
    Radius-velocity and position-position distributions for our \wx\
    field, as described in \ref{fig_phase.w05}.}
\label{fig_phase.w10}
\end{figure}

As a consistency check on our member/non-member division in velocity
space, we have estimated the cluster escape velocity as a function of
projected clustocentric radius and displayed the corresponding loci as
lines in panels (a) and (c) of Figures~\ref{fig_phase.w05} through
\ref{fig_phase.w10}.  The estimates assume isotropic orbits and a
\cite{her90} density profile given by

\begin{equation}
  \rho = \frac{M_{cl}}{2\pi R_c R (1+R/R_c)^{3}}
\end{equation}
where $M_{cl}$ is the total cluster mass and $R_c$ is the core radius,
normalized, respectively to $M_{200}$ and $\Rvirial$ from
Table~\ref{tab_velocity_distrib_hist} here and in \cite{cra11}. \cite{car97} found that the relation $R_c/\Rvirial =
0.66\pm0.09$ yielded the best fit to the galaxy surface density
ensemble for 16 clusters in the redshift range $0.17<z<0.55$,
including two of our clusters (\woi\ and \wov); we adopt their radial
size scaling.  However, their estimate of $\sigma$ for \woi\ is
smaller than ours by about $16\%$.  We attribute this to two factors:
(1)~rather than attempting to distinguish individual cluster members
from fore- and background galaxies, Carlberg \etal\ employed a
statistical correction which is subject to error; (2)~we have the
benefit of a substantially larger, representative sample of
spectroscopic redshifts within $\Rvirial$.  If we re-derive an estimate
of $\sigma$ using only the Carlberg \etal\ sample, we find results
consistent with theirs.  The higher value of $\sigma$ we obtain is
consistent with recent theoretical work \citep{old13, Wu13} indicating
that a large sample of galaxies randomly selected over a range of
magnitude and radius is needed to accurately measure the velocity
dispersion of a cluster.

A comparison of our measured velocity dispersions with those from
other studies shows good general agreement.  Our measured value of the
velocity dispersion of \wov\ is in excellent agreement with
\cite{borgani99}, who measure a value of $\sigma = 1330^{+111}_{-94}$~\kms.  \cite{jorgensen13} measure a slightly higher value of $\sigma=
1450^{+105}_{-159}$~\kms\ based on only 47 sources.  Our value for
\wovii\ is lower, but still in agreement, with the velocity
dispersion of $\sigma = 1156\pm82$~\kms\ from \cite{tran07}.  The
results for \woviii\ and \wx\ are both smaller than the
previous measuments by \cite{lubin04} of $\sigma = 1016^{+126}_{-96} $
\kms\ and $\sigma = 1226^{+245}_{-154}$ \kms, respectively.  Our
result for \woviii\ is significantly smaller than the previous
measurements, whereas the difference for \wx\ is not.

A range of line-of-sight escape velocities exists at any given
projected radius. Within the uncertainties of the adopted values of
$R_c/\Rvirial$ and these projection effects, the identified cluster
galaxies do indeed lie within a region of phase space consistent with
being \emph{bona fide} cluster members. Detailed mass modeling of, and
galaxy orbits within, the cluster is beyond the scope of this
work. Here, our primary aim is to derive the characteristic values of
the average line-of-sight velocity distribution and spatial projection
of cluster galaxies to determine whether they differ by type.

\subsection{Projected Velocity Distribution}
\label{sxn-velocity}

\begin{figure}[p]
\plotone{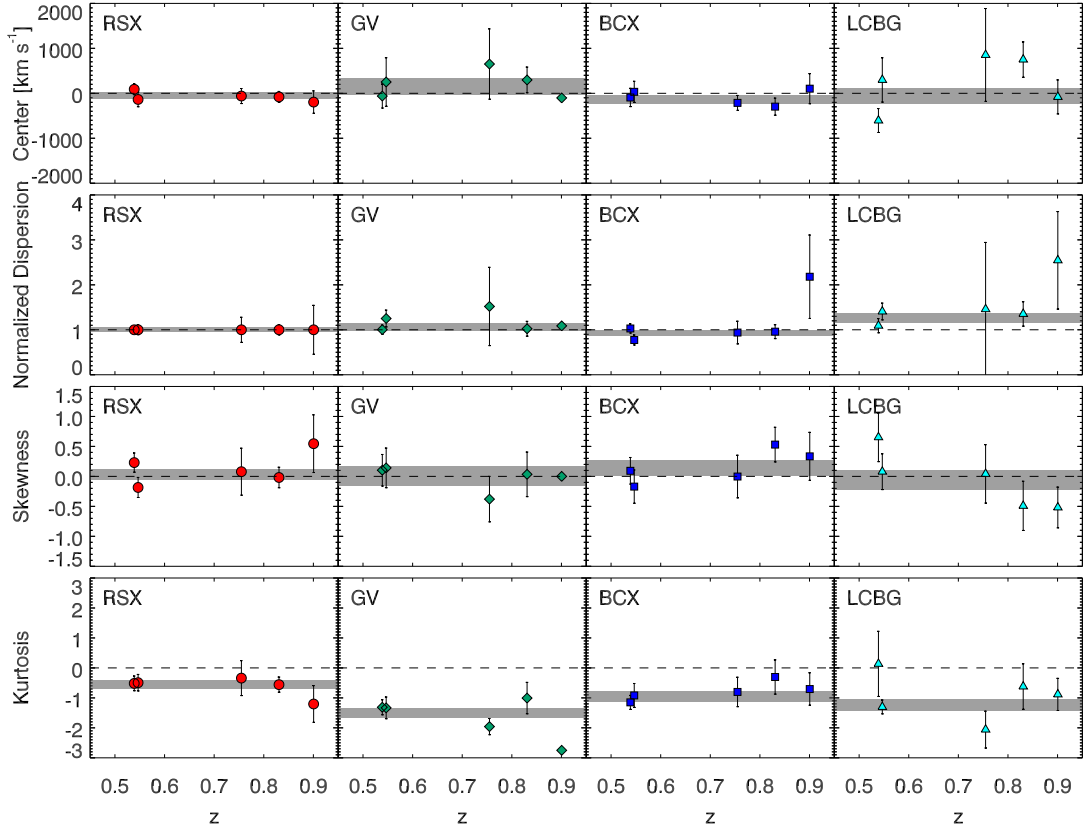}
\caption{ Measures of velocity offset, dispersion, skewness, and
  kurtosis for the galaxy classes in each of our cluster fields.
  Panels show the relevant measurements (and corresponding errors
  derived from bootstrap resampling) in our five clusters for a
  particular galaxy type as indicated on the plot.  Panels in the top
  row depict offset from the nominal cluster redshift; the dashed line
  indicates zero offset.  Panels in the second row show the ratio of
  $\veldisp$ measurements for each galaxy class in each of our cluster
  fields vs.\ the value of $\veldisp$ for the RSX-class objects in
  that cluster.  The horizontal dashed line at a value of 1 indicates
  equality of the scale measurements for the indicated subclass and
  the RSX subclass.  Panels in the third row show skewness.  Positive
  skewness indicates that the distribution is skewed to velocities
  greater than the cluster mean.  Panels in the bottom row display
  kurtosis.  Positive kurtosis indicates that the distribution has
  more weight in the tail than a Gaussian.  In each panel, the shaded
  region is centered on the weighted mean value of the clusters in the
  sample and spans the $\pm1~\sigma$ confidence interval on the mean.}
\label{fig_cluster_stats4}
\end{figure}

To determine quantitatively whether the radial velocity distributions
for the various subpopulations differ, we computed the first four
moments of the differential velocity distribution (offset, dispersion,
skewness, and kurtosis); results appear in
Table~\ref{tab_velocity_distrib_hist} and
Figure~\ref{fig_cluster_stats4}.

%----------------------------------------
% tab_velocity_distrib_hist
%----------------------------------------
\begin{deluxetable}{llrr@{}c@{}lr@{}c@{}lr@{}c@{}lr@{}c@{}l}
\tabletypesize{\footnotesize}
  \tablewidth{0pc}
  \tablecaption{Velocity Distribution Parameters for Individual Clusters
    \label{tab_velocity_distrib_hist}
  }
  \tablehead{
    \colhead{Cluster} &
    \colhead{Type} &
    \colhead{$N_g$}  &
    \multicolumn{3}{c}{Offset}         &
    \multicolumn{3}{c}{Dispersion} &
    \multicolumn{3}{c}{Skewness} &
    \multicolumn{3}{c}{Kurtosis}  \\
    &
    &
    &
    \multicolumn{3}{c}{(\kms)} &
    \multicolumn{3}{c}{(\kms)} &
    \multicolumn{3}{c}{} &
    \multicolumn{3}{c}{}  \\
    \colhead{(1)} &
    \colhead{(2)} &
    \colhead{(3)} &
    \multicolumn{3}{c}{(4)} &
    \multicolumn{3}{c}{(5)} &
    \multicolumn{3}{c}{(6)} &
    \multicolumn{3}{c}{(7)}
  }
  \startdata
  %% filename = tab_velocity_distrib_hist.dat
%% creator  = tab_velocity_distrib_hist.pro
%% date     = Wed Sep 25 21:36:06 2013
%% column1  = cluster
%% column2  = subset
%% column3  = N_members
%% column4  = delta_v [km/s]
%% column5  = scale
%% column6  = skewness
%% column7  = kurtosis
\wov & RSX & 118 & $91$ & $\pm$ & $122$ & $1262$ & $\pm$ & $72$ & $0.2$ & $\pm$ & $0.2$ & $ -0.5$ & $\pm$ & $0.2$ \\
\wov & GV & 26 & $-59$ & $\pm$ & $266$ & $1266$ & $\pm$ & $126$ & $0.1$ & $\pm$ & $0.3$ & $ -1.3$ & $\pm$ & $0.3$ \\
\wov & BCX & 44 & $-91$ & $\pm$ & $203$ & $1299$ & $\pm$ & $103$ & $0.1$ & $\pm$ & $0.2$ & $ -1.1$ & $\pm$ & $0.2$ \\
\wov & LCBG & 32 & $-604$ & $\pm$ & $266$ & $1376$ & $\pm$ & $182$ & $0.7$ & $\pm$ & $0.4$ & $ 0.1$ & $\pm$ & $1.1$ \\
~\\
\woi & RSX & 92 & $-135$ & $\pm$ & $162$ & $1462$ & $\pm$ & $113$ & $-0.2$ & $\pm$ & $0.2$ & $ -0.5$ & $\pm$ & $0.3$ \\
\woi & GV & 16 & $254$ & $\pm$ & $540$ & $1829$ & $\pm$ & $272$ & $0.1$ & $\pm$ & $0.3$ & $ -1.3$ & $\pm$ & $0.4$ \\
\woi & BCX & 28 & $33$ & $\pm$ & $234$ & $1130$ & $\pm$ & $136$ & $-0.2$ & $\pm$ & $0.3$ & $ -0.9$ & $\pm$ & $0.4$ \\
\woi & LCBG & 24 & $303$ & $\pm$ & $493$ & $2057$ & $\pm$ & $220$ & $0.1$ & $\pm$ & $0.3$ & $ -1.3$ & $\pm$ & $0.2$ \\
~\\
\woviii & RSX & 20 & $-60$ & $\pm$ & $166$ & $682$ & $\pm$ & $134$ & $0.1$ & $\pm$ & $0.4$ & $ -0.3$ & $\pm$ & $0.6$ \\
\woviii & GV & 4 & $651$ & $\pm$ & $781$ & $1037$ & $\pm$ & $594$ & $-0.4$ & $\pm$ & $0.4$ & $ -2.0$ & $\pm$ & $0.3$ \\
\woviii & BCX & 17 & $-212$ & $\pm$ & $164$ & $641$ & $\pm$ & $116$ & $-0.0$ & $\pm$ & $0.4$ & $ -0.8$ & $\pm$ & $0.5$ \\
\woviii & LCBG & 5 & $854$ & $\pm$ & $1032$ & $995$ & $\pm$ & $994$ & $0.0$ & $\pm$ & $0.5$ & $ -2.1$ & $\pm$ & $0.6$ \\
~\\
\wovii & RSX & 80 & $-79$ & $\pm$ & $122$ & $1039$ & $\pm$ & $83$ & $-0.0$ & $\pm$ & $0.2$ & $ -0.6$ & $\pm$ & $0.3$ \\
\wovii & GV & 18 & $297$ & $\pm$ & $285$ & $1063$ & $\pm$ & $170$ & $0.0$ & $\pm$ & $0.4$ & $ -1.0$ & $\pm$ & $0.5$ \\
\wovii & BCX & 37 & $-297$ & $\pm$ & $189$ & $996$ & $\pm$ & $139$ & $0.5$ & $\pm$ & $0.3$ & $ -0.3$ & $\pm$ & $0.6$ \\
\wovii & LCBG & 16 & $751$ & $\pm$ & $392$ & $1407$ & $\pm$ & $258$ & $-0.5$ & $\pm$ & $0.4$ & $ -0.6$ & $\pm$ & $0.8$ \\
~\\
\wx & RSX & 7 & $-195$ & $\pm$ & $249$ & $524$ & $\pm$ & $201$ & $0.5$ & $\pm$ & $0.5$ & $ -1.2$ & $\pm$ & $0.6$ \\
\wx & GV & 2 & $-102$ & $\pm$ & $\cdots$ & $570$ & $\pm$ & $\cdots$ & $0.0$ & $\pm$ & $\cdots$ & $ -2.8$ & $\pm$ & $\cdots$ \\
\wx & BCX & 17 & $103$ & $\pm$ & $334$ & $1143$ & $\pm$ & $208$ & $0.3$ & $\pm$ & $0.4$ & $ -0.7$ & $\pm$ & $0.5$ \\
\wx & LCBG & 16 & $-78$ & $\pm$ & $377$ & $1335$ & $\pm$ & $245$ & $-0.5$ & $\pm$ & $0.3$ & $ -0.9$ & $\pm$ & $0.5$ \\

  \enddata
  \tablecomments{(1)~Cluster name;
    (2)~galaxy class as described in text; (3)~number of galaxies in
    subset; (4)~velocity offset of this subset from the full sample;
    (5)~velocity scale of the subset; (6)~skewness of the subset;
    (7)~kurtosis of the subset
  }
\end{deluxetable}

Our analysis reveals no compelling evidence that the first and third
moments (barycenter and velocity skew) differ consistently between the
cluster subpopulations. However, we do detect certain differences in
the even moments between various subsamples.  Specifically, we find
that the velocity distribution is generally platykurtic (\ie,
displaying negative kurtosis; boxier than Gaussian).  This effect
appears to be most pronounced for the GV subsample and somewhat
larger for the bluer, star-forming populations relative to the red
population overall.  

In terms of dispersion (the second moment), it is
easiest to compare the populations if we normalize the dispersions
within a given cluster to the dispersion for that cluster's dominant
RSX population, as shown in Fig.~\ref{fig_cluster_stats4} and
Table~\ref{tab_velocity_distrib_hist}. Here, we observe that the LCBG
population has a velocity dispersion which exceeds that of the RCX
sample by a factor of $1.56\pm0.5$, possibly increasing with redshift.
A comparison between the radial velocity distributions for the LCBGs
and red cluster members via the Komolgorov-Smirnov (K-S) statistical
test indicates that we can reject with $>90\%$ confidence the
hypothesis that LCBGs are drawn from the same parent population as the
RSX galaxies. For the GV and BCX classes, the evidence is mixed.
Interpretation of the larger LCBG velocity dispersion depends on their
relative spatial distribution, but certainly one plausible explanation
is that this population is not yet virialized.  The expected ratio
between the velocity dispersion of a virialized population and an
infalling population is $\sqrt{2}$ \citep{cd96, biviano97}, which is
consistent with the value found for the LCGGs here.  In short, the
GV and LCBG classes appear to stand out in their velocity
distributions as being, respectively, boxier and broader.  We return
to the implications of these findings in \S\ref{discussion}.

\subsection{Projected Spatial Distribution}
\label{sxn-space}

\begin{figure}[p]
\epsscale{0.9}
  \plotone{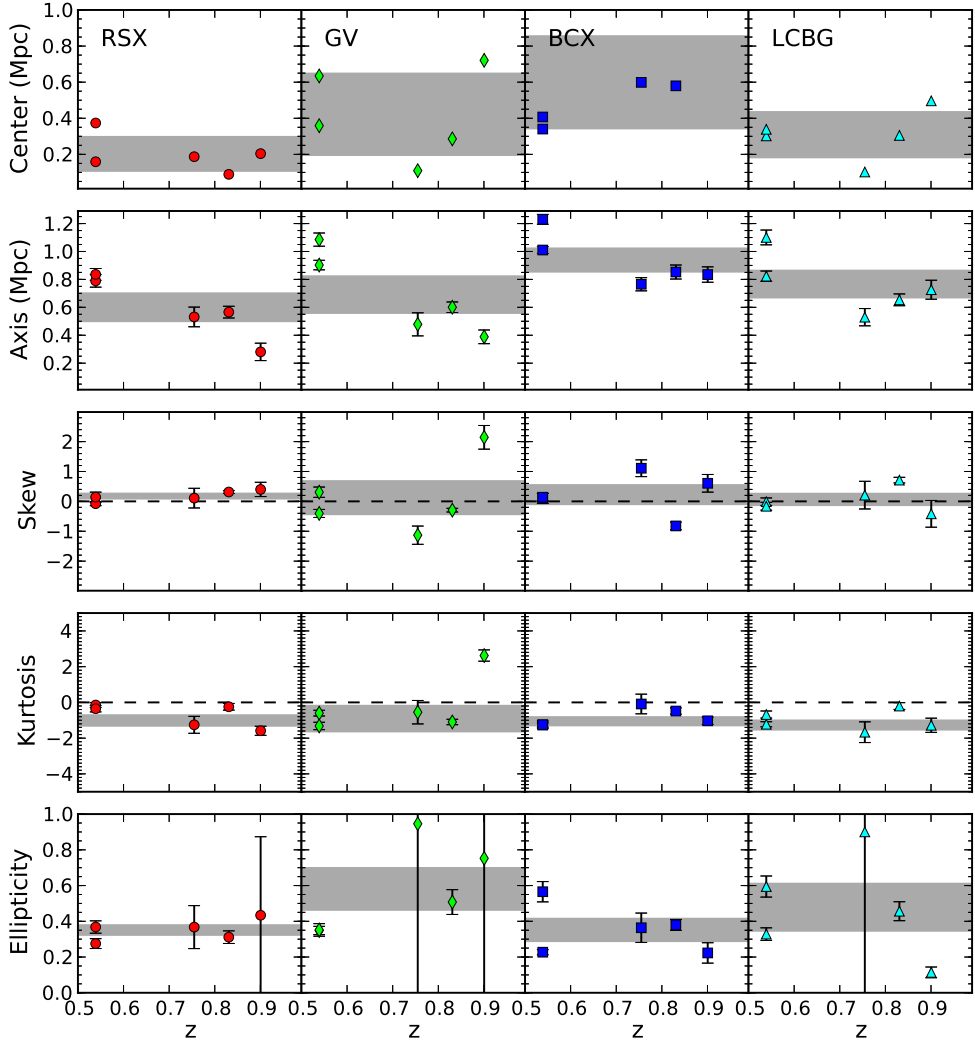}
  \caption{
    \label{fig_spatial_stats}
    Spatial statistics for cluster sources (top to bottom): offset of
    the galaxy cluster centroid from the BCG; major axis length (as defined
    in text); skew along the major axis; kurtoisis along the major
    axis; and ellipticity.  Data are plotted for the following
    galaxies samples respectively: all, red sequence, blue cloud,
    green valley, and LCBG.  }
\end{figure}

The spatial distributions of the different cluster populations are
presented in Figures \ref{fig_phase.w01}--\ref{fig_phase.w10}.  For
each cluster and galaxy class, we also calculate the center, major
axis length $a$ (\ie, the second moment of the galaxy spatial
distribution), ellipticity $e$, position angle $\theta$, skewness, and
kurtosis of the distribution. To compensate for incompleteness in our
sample, we applied a correction derived by estimating the completeness
of the spectroscopy as a function of projected position and spectral
type; each object is then weighted by the likelihood of having a
measured spectroscopic redshift.  The weight is calculated both based
on spectroscopic sampling and color of the object.  The grayscale
background in Figures \ref{fig_phase.w05}--\ref{fig_phase.w10}(d)
represent the overall completeness as a function of spatial position
for each of our fields. We present the spatial distribution statistics
in Figure~\ref{fig_spatial_stats} and Table~\ref{tab_spatial_params}.

%----------------------------------------
% tab_spatial_params
%----------------------------------------
\begin{deluxetable}{llrrrrrrrrr}
\tabletypesize{\footnotesize}
  \tablewidth{0pc}
  \tablecaption{Spatial Distribution Parameters for
    Individual Clusters
    \label{tab_spatial_params}
  }
  \tablehead{
    \colhead{Cluster} &
    \colhead{Type} &
    \colhead{$\alpha$}  &
    \colhead{$\delta$} &
    \colhead{$a$}  &
    \colhead{Skewness} &
    \colhead{Kurtosis} &
    \colhead{$e$}  &
    \colhead{$\theta$}  \\
    \colhead{} &
    \colhead{} &
    \colhead{(J2000)} &
    \colhead{(J2000)} &
    \colhead{(Mpc)} &
    \colhead{} &
    \colhead{} &
    \colhead{} &
    \colhead{($\deg$)} \\
    \colhead{(1)} &
    \colhead{(2)} &
    \colhead{(3)} &
    \colhead{(4)} &
    \colhead{(5)} &
    \colhead{(6)} &
    \colhead{(7)} &
    \colhead{(8)} &
    \colhead{(9)}
  }
  \startdata
  \woi\ &   ALL &  73.55095 & -3.01563 &  $0.91\pm0.03$ & $-0.65\pm0.15$ & $0.96\pm0.60$ & $0.34\pm0.01$ & $-42.8\pm3.8$ \\ 
\woi\ &   RSX &  73.55041 & -3.00985 &  $0.79\pm0.05$ & $-0.15\pm0.15$ & $1.85\pm0.53$ & $0.37\pm0.01$ & $-37.3\pm3.7$ \\ 
\woi\ &    GV &  73.56060 & -3.01321 &  $1.09\pm0.04$ & $-0.60\pm0.17$ & $0.36\pm0.60$ & $0.35\pm0.01$ & $-43.5\pm2.9$ \\ 
\woi\ &    BCX &  73.54606 & -3.03185 &  $1.01\pm0.05$ & $-1.27\pm0.12$ & $-0.79\pm0.41$ & $0.23\pm0.01$ & $-25.5\pm3.9$ \\ 
\woi\ &  LCBG &  73.55579 & -3.02170 &  $1.10\pm0.05$ & $-1.22\pm0.25$ & $-0.58\pm0.71$ & $0.59\pm0.01$ & $-42.7\pm4.1$ \\ 
\\
\wov\ &   ALL &   4.63501 & 16.42686 &  $0.94\pm0.06$ & $-0.71\pm0.17$ & $0.49\pm0.51$ & $0.36\pm0.01$ & $39.9\pm4.3$ \\ 
\wov\ &   RSX &   4.63379 & 16.42283 &  $0.84\pm0.04$ & $-0.35\pm0.21$ & $1.34\pm0.89$ & $0.28\pm0.01$ & $43.0\pm4.2$ \\ 
\wov\ &    GV &   4.62705 & 16.41352 &  $0.90\pm0.05$ & $-1.32\pm0.18$ & $-1.42\pm0.61$ & $0.35\pm0.01$ & $31.2\pm4.2$ \\ 
\wov\ &    BCX &   4.62593 & 16.43336 &  $1.23\pm0.04$ & $-1.25\pm0.22$ & $-0.29\pm0.49$ & $0.57\pm0.01$ & $35.2\pm4.4$ \\ 
\wov\ &  LCBG &   4.65458 & 16.43949 &  $0.82\pm0.05$ & $-0.68\pm0.16$ & $-0.60\pm0.57$ & $0.33\pm0.01$ & $16.3\pm4.4$ \\ 
\\
\woviii\ &   ALL &  201.20370 & 30.18699 &  $0.57\pm0.06$ & $-0.49\pm0.51$ & $0.60\pm0.98$ & $0.23\pm0.01$ & $19.9\pm17.6$ \\ 
\woviii\ &   RSX &  201.20979 & 30.19631 &  $0.53\pm0.05$ & $-1.26\pm0.60$ & $-0.72\pm0.86$ & $0.37\pm0.01$ & $ 1.5\pm13.5$ \\ 
\woviii\ &    GV &  201.20027 & 30.19101 &  $0.48\pm0.04$ & $-0.55\pm0.53$ & $-0.43\pm0.94$ & $0.95\pm0.01$ & $-17.1\pm12.0$ \\ 
\woviii\ &    BCX &  201.19697 & 30.17219 &  $0.77\pm0.07$ & $-0.09\pm0.45$ & $-1.10\pm0.99$ & $0.36\pm0.01$ & $10.1\pm22.2$ \\ 
\woviii\ &  LCBG &  201.20547 & 30.19718 &  $0.53\pm0.09$ & $-1.67\pm0.55$ & $-1.76\pm0.95$ & $0.90\pm0.01$ & $34.4\pm21.9$ \\ 
\\
\wovii\ &   ALL &  164.25368 & -3.62277 &  $0.71\pm0.03$ & $-0.78\pm0.12$ & $0.43\pm0.46$ & $0.39\pm0.01$ & $26.3\pm4.3$ \\ 
\wovii\ &   RSX &  164.24792 & -3.62918 &  $0.57\pm0.05$ & $-0.25\pm0.23$ & $1.79\pm0.51$ & $0.31\pm0.01$ & $22.6\pm4.4$ \\ 
\wovii\ &    GV &  164.23990 & -3.62933 &  $0.60\pm0.04$ & $-1.10\pm0.11$ & $-0.70\pm0.59$ & $0.51\pm0.01$ & $29.3\pm5.2$ \\ 
\wovii\ &    BCX &  164.26722 & -3.61432 &  $0.85\pm0.04$ & $-0.48\pm0.19$ & $-0.42\pm0.53$ & $0.38\pm0.01$ & $18.3\pm4.7$ \\ 
\wovii\ &  LCBG &  164.25282 & -3.61589 &  $0.65\pm0.02$ & $-0.20\pm0.12$ & $-0.39\pm0.54$ & $0.46\pm0.01$ & $33.7\pm4.5$ \\ 
\\
\wx\ &   ALL &  241.07581 & 43.08096 &  $0.69\pm0.08$ & $-1.01\pm1.14$ & $-0.75\pm0.60$ & $0.04\pm0.01$ & $42.6\pm39.2$ \\ 
\wx\ &   RSX &  241.09303 & 43.07960 &  $0.28\pm0.07$ & $-1.59\pm0.37$ & $-0.95\pm0.38$ & $0.43\pm0.01$ & $-25.7\pm39.5$ \\ 
\wx\ &    GV &  241.08517 & 43.09851 &  $0.39\pm0.08$ & $2.62\pm0.27$ & $-1.69\pm0.59$ & $0.75\pm0.01$ & $-39.9\pm58.4$ \\ 
\wx\ &    BCX &  241.06191 & 43.08007 &  $0.83\pm0.08$ & $-1.03\pm0.26$ & $-1.63\pm0.45$ & $0.22\pm0.01$ & $-4.6\pm31.9$ \\ 
\wx\ &  LCBG &  241.08262 & 43.08087 &  $0.73\pm0.05$ & $-1.28\pm1.34$ & $-1.30\pm0.56$ & $0.11\pm0.01$ & $44.4\pm35.2$ \\ 
\\

  \enddata
  \tablecomments{(1)~Cluster name;
    (2)~Galaxy class as described in text; (3)~RA center of subset;
    (4)~Declination of subset; (5)~Major axis of subset; (6)~Kurtosis along major axis
     of subset; (7)~Kurtosis along major axis of subset;
    (8)~Ellipticity of distribution; (9)~Position angle of subset on sky.
  }
\end{deluxetable}

The analyis reveals similar spatial-distribution characteristics
(ellipticity, skew, kurtosis) between subpopulations in a given
cluster and between clusters as a whole. There is little skew, mild,
negative kurtosis ($>-1$), and a mean cluster ellipticity of
$\sim$0.4. The greatest distinction between subpopulations is
exhibited in the size distribution and de-center relative to the
BCG. The BCX population has the largest spatial distribution and the
largest offset on average, while the RCX values are the smallest. The
BCX decenter values are three times more than that for the RCX
galaxies, but still well below the cluster core radius. RCX size
distribution is roughly 60\% that for the BCX population, or roughly
20\% of the volume.  The GV and LCBG values are intermediate between
the BCX and RCX population values in the mean, but exhibit more
scatter from cluster to cluster.

\subsubsection{Projected Radial Distribution}

\begin{figure}[p]
\epsscale{0.9}
  \plotone{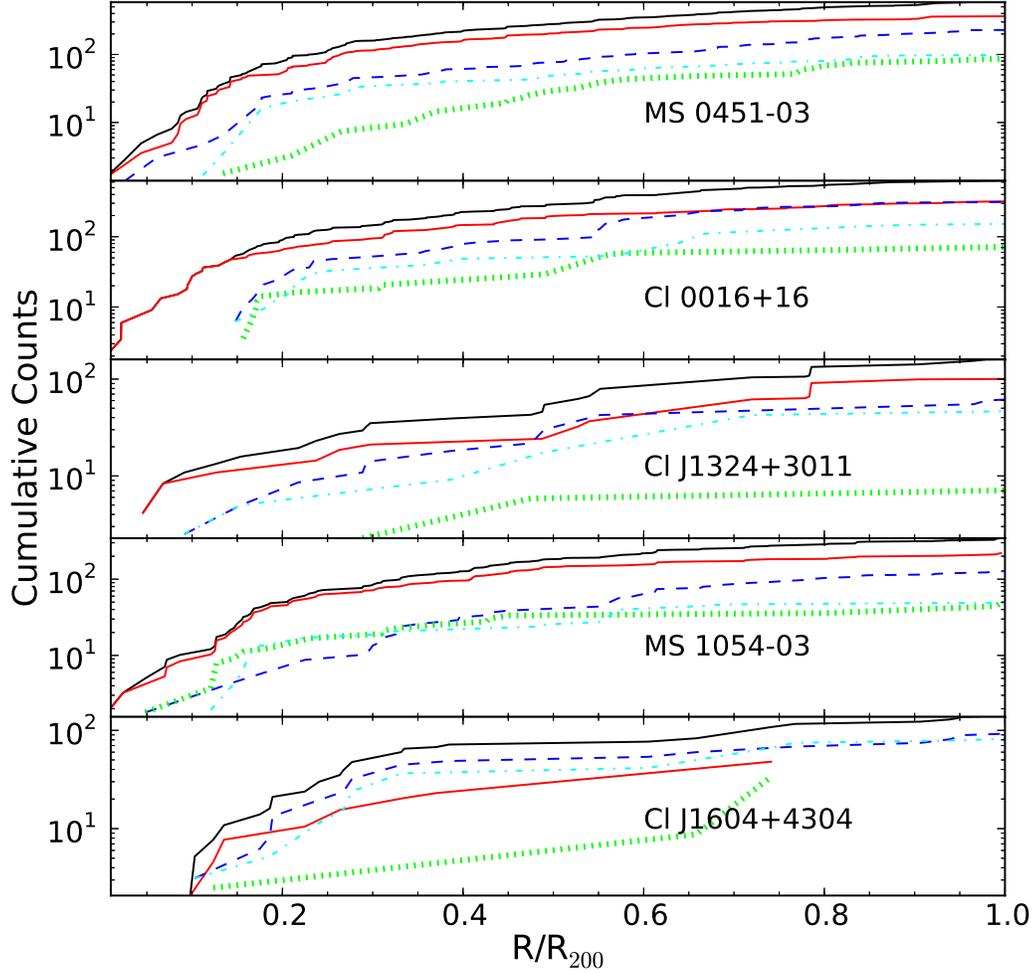}
  \caption{
    \label{fig_radhist}
    Radial, cumulative histograms for the spatial distributions for
    different spectral types along the cluster major-axis: all types
    (black solid), RSX class (red solid), BCX class (blue dashed), GV class (green dotted), and
    LCBG class (turquoise dot-dashed). Histograms have been corrected for cluster
    ellipticity and sampling completeness, and are calculated from the
    source-distribution center (\ie, not the location of the BCG).  }
\end{figure}

We also computed the projected radial surface density distributions
for each of the different populations and present the results in
Figure~\ref{fig_radhist}.  To account for spectroscopic incompleteness
in our survey, we applied a correction in computing these
distributions following \cite{cra11}.  The first step was to calculate
the photometric galaxy classification assuming every object in the
field was at the redshift of the cluster.  We then summed the number
of galaxies in a given classification within each of our radial bins.
Next, we computed the ratio of spectroscopically-confirmed cluster
members of a given class to the total number of objects with
spectroscopy that had the same photometric galaxy classification.
This ratio was calculated within each of the radial bins that we used
and then applied to that radial bin.  Finally, the corrected number
density of galaxies was divided by the area of the radial bin to
produce the surface density of galaxies.  We only measured the surface
density profiles out to the terminal radius of our WIYN survey.

\subsubsection{Trends and characteristics}

All of these clusters exhibit a strong red sequence that peaks in the
central region and falls off with radius, while the other classes of
galaxies tend to show ``cored'' profiles (\ie, profiles with a deficit
of galaxies in the central region). This contrast can be seen in
Fig.~\ref{fig_radhist} as a paucity of all but RCX galaxies within
$0.2\Rvirial$, particularly in the three most massive clusters.  For
BC, LCBG and GV types, the number of galaxies with $R/\Rvirial<0.2$ is
significantly less than the number of galaxies that would be expected
if they had the same profile as red sequence galaxies.  This relative
lack of star-forming galaxies observed in the cluster core appears
consistent with previous studies of rich clusters \citep{thompson86,
  ellingson01,cra06,mahajan10}.

Within the three higher-redshift clusters, the spatial distribution of
member galaxies appears more irregular.  The most massive of the
three, \wov, shows similar behavior to the two lower-redshift
clusters, but with a much higher density of GV and BC objects within
its core.  The two slightly-lower-mass clusters show a much more
irregular distribution of member galaxies.  This pair of clusters
shows far more blue galaxies within their central regions.  This
diminution of the morphology-density relationship has also been
reported previously \citep[\eg,][]{cooper08, tran10, hilton10,
  hayashi10}.

Visual inspection of the clusters reveals distinct patterns in the
spatial distribution of the galaxies of different spectral type, which
we characterize here:

\begin{description}
  
\item[\wov] (Fig.~\ref{fig_phase.w05}) appears as a concentrated,
  massive cluster with the central region dominated by RS galaxies.
  Its shape is an elongated ellipse with higher galaxy density along
  the major axis, an effect previously reported by \cite{moran07}.
  Most of the red galaxies lie along this axis, whereas blue galaxies
  are primarily located on the periphery.  Moran \etal\ posited that
  \wov\ had two large filaments feeding it, which explains the spatial
  and velocity distribution seen for the cluster.  A very high
  fraction of blue galaxies (56\%) on the interior of the cluster are
  LCBGs, while only 9\% of galaxies within the core of the cluster
  ($R< 0.5\Rvirial$) are GV galaxies.  
  
\item[\woi] (Fig.~\ref{fig_phase.w01}) is a massive galaxy cluster
  dominated by a very strong and extended distribution of RS galaxies.
  These galaxies form a highly elliptical distribution with evidence
  of strong filamentary structure beyond $\Rvirial$.  Similar to \wov,
  43\% of blue galaxies are LCBGs, but such galaxies are nearly absent
  in the cluster core.  GV galaxies are distributed throughout the
  cluster, but still comprise only 12\% of the population within
  $\Rvirial$.
 
\item[\woviii] (Fig.~\ref{fig_phase.w08}) is a very diffuse cluster
  with a small population of RS galaxies.  Blue galaxies are the
  dominant subclass within the cluster, although RS galaxies form a
  small core which is offset from the overall distribution.  The ratio
  of LCBGs to blue galaxies is relatively high at 50\%, whereas GV
  galaxies comprise 13\% of the total population.
  
\item[\wovii] (Fig.~\ref{fig_phase.w07}) is a higher-redshift analogue
  to \wov\ and \woi, with a compact core of RS galaxies, an overall
  elliptical galaxy distribution, and blue galaxies dominating the
  periphery.  It has a high fraction of blue galaxies (56\%) within
  the cluster core.  GV galaxies are spread throughout the cluster
  and have a similar distribution to the blue galaxies, comprising
  18\% of the population.
  
\item[\wx] (Fig.~\ref{fig_phase.w10}) is also a very diffuse cluster
  with an extended distribution of blue galaxies around a loose core
  of RS glaxies.  The ratio of LCBGs to blue galaxies is extreme at
  66\%, but the cluster is nearly devoid of GV galaxies (6\%
  of all sources).

\end{description}

\section{Subclustering}

\subsection{Dressler-Shectman Test}

We applied the Dressler-Shectman (DS) statistical test to determine
whether significant substructure exists in our clusters \citep{ds88}.
The classical DS statistic is calculated by computing a local velocity
mean ($\bar{v}_{local} $) and velocity dispersion ($\sigma_{local}$)
for each galaxy and comparing these to the global values for the
cluster.  It is calculated via the relation
\begin{equation}
\delta^2 = \frac{N_{obj}}{\sigma^2}[(\bar{v}_{local}-\bar{v})^2
+((\sigma_{local}-\sigma)^2]
\end{equation}
where $N_{obj}$, the number of nearest neighbors considered in the
computation, is traditionally taken to be 10.  As such, the statistic
measures substructure in bulk flow and velocity dispersion on a
density-dependent spatial scale. For each cluster, we also compute
$\Delta = \sum \delta$, a quantity which charcterizes the overall
substructuring in a cluster.  The significance of the
substructure can be determined by comparing $\Delta$ to the results of
a Monte Carlo simulation, where the velocities of the sources have
been randomized.  Each simulation was run 1000 times to compute the
probability $P$ that $\Delta_{sim} > \Delta_{obs}$.  We have computed
$\Delta$ for the overall galaxy population, but also for each of our
subclasses.  For each of the tests, we only measured $\delta$ for
galaxies that lie within the WIYN field of view so that we could
compare the different subclasses.

\begin{figure}[p]
\epsscale{0.9}
  \plotone{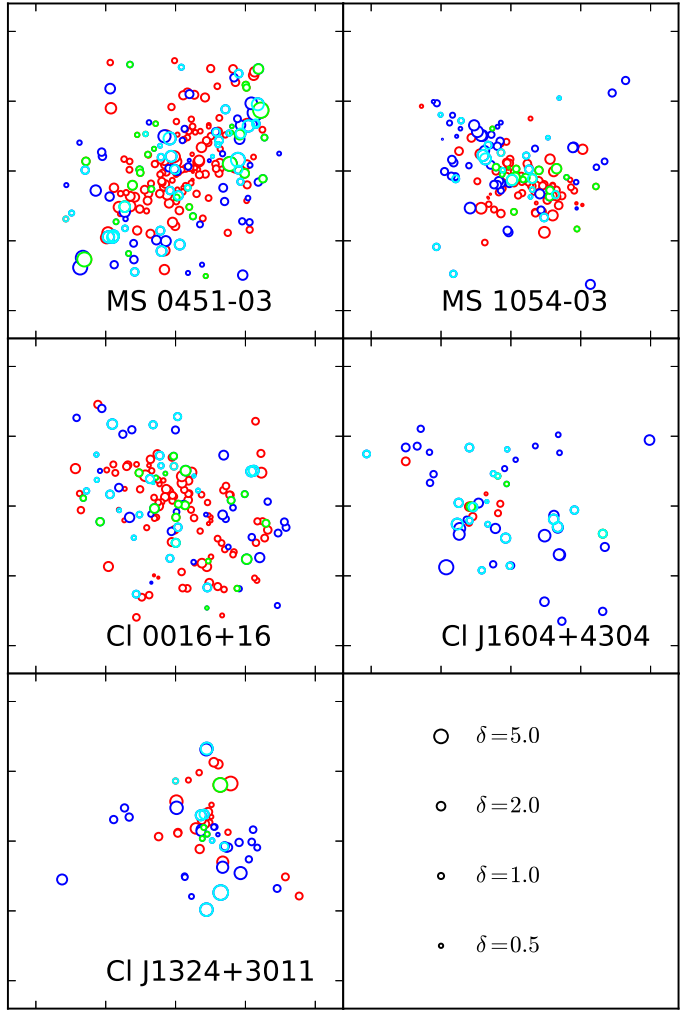}
  \caption{
    \label{fig_dsall}   
    Dressler-Shectman (DS) measurements for substructure in each of
    our clusters. Each circle represents a bonafide cluster member,
    size-coded by the DS statistics and color-coded by spectral type:
    RCX (red), BCX (blue)), GV (green), and LCBG (turquoise). The size
    scale for the DS statistic is given in the bottom right panel
    }
\end{figure}

The results for each of our clusters appear in Fig.~\ref{fig_dsall}.
Individual clusters are described in detail below:

\begin{description}
  
\item[\wov] shows signs for some substructure as compared to the
  simulations with only 7\% of the simulations having $\Delta$ as high
  as that observed for the entire cluster population.  The elongated
  structure has been reported previously \citep{don03,moran07}, and a
  triaxial distribution to the gas has been postulated as the most
  consistent fit to the observations of the gas distribution
  \citep{don03}.  In this cluster, the blue populations show more
  signal for subclustering than the red populations with the LCBG
  class having the highest indication of subclustering.
  
\item[\woi] shows no signs for substructure, with a very smooth,
  uniform distribution of galaxies. The cluster does exhibit a
  slightly elongated structure and is embedded in a much larger
  superstructure \citep{tanaka2007}.  None of the subpopulations show
  any significant differences in subclustering.
  
\item[\woviii] shows no signs for substructure.  However, the small
  population of LCBGs in this cluster does show a higher degree of
  clumping than the other populations and there is a fairly high
  degree of clustering within each of the populations.
  
\item[\wovii] does show significant evidence for substructure with
  only 5\% of the simulations showing the same degree of substructure.
  There is also an elongated structure to the cluster as well.
  However, none of the subpopulations show significantly more
  substructure than the others.
  
\item[\wx] does not show any significant substructure.  The
  distribution of objects in the cluster is very diffuse.  The blue
  galaxies and the LCBGs do show evidence for being more clustered
  into substructures than the other populations.

\end{description}

On the whole, the clusters present no coherent picture of the
association of blue galaxies with substructure on the scale of 10
nearest neighbors.  For the three most massive clusters, only a small
fraction of the RS galaxies were associated with substructure, and
only in \wov\ was a large number of blue galaxies associated with
substructure.

\subsection{Subclustering on different scales}

\begin{figure}[p]
\epsscale{0.8}
  \plotone{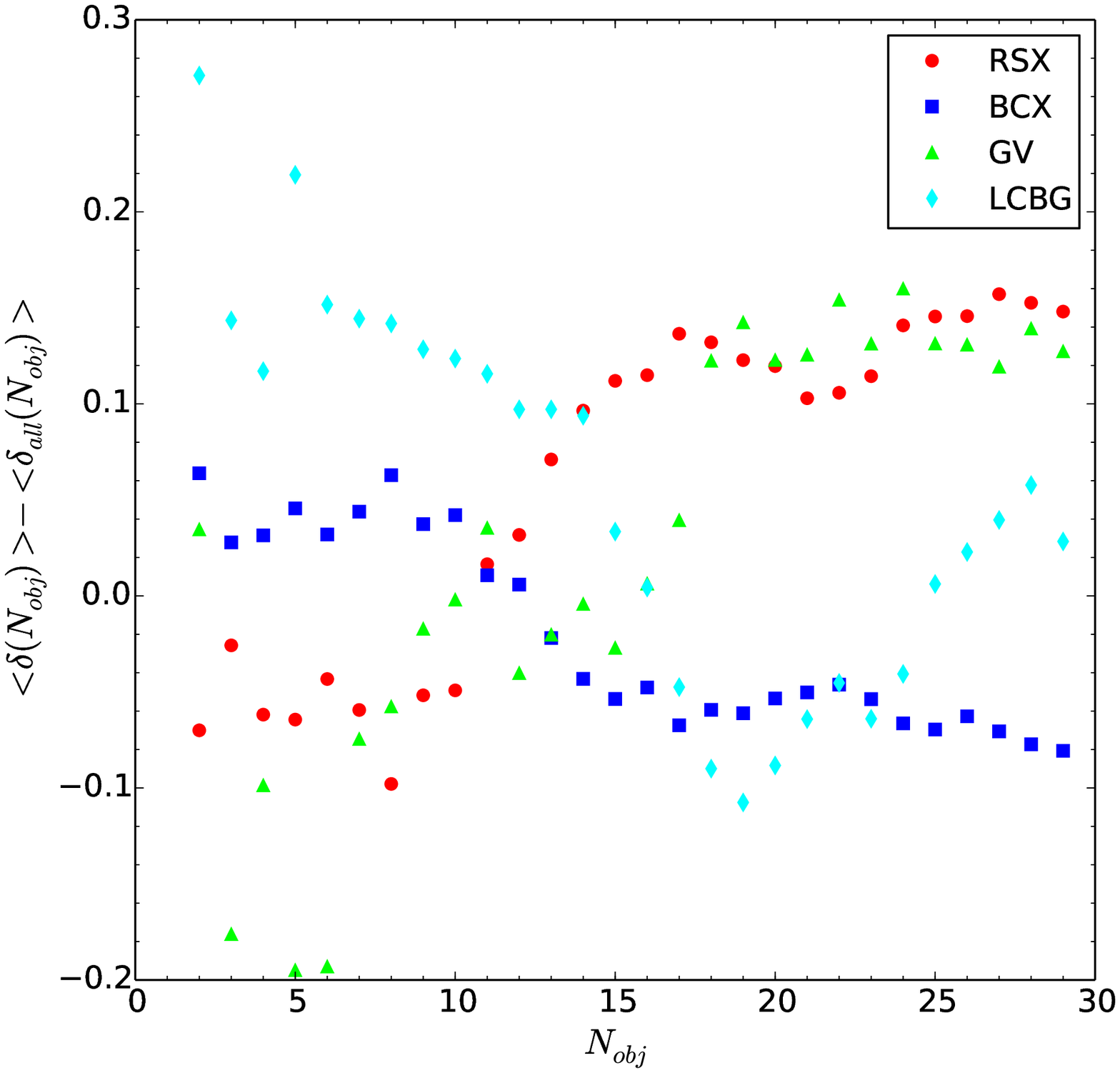}
  \caption{
   \label{fig:dsave}
   Offset in DS statistic for each of our different
   populations for different values of $N_{obj}$, the number of
   neighbors, from the DS value for the cluster.  This statistic has
   been averaged over all clusters.   In the figure, RSX are
   representated by red circles; BCX, blue squares; GV, green
   triangles, and LCBGs, teal diamonds. }
\end{figure}

For each subclass of galaxy, we investigated how the galaxies were
distributed on different (density-dependent spatial) scales by varying
the number of nearest neighbors to seek trends in subclustering.  We
made repeated measurements of the DS statistic with values of
$N_{obj}$ ranging from 2 to 30, following \cite{cd96}.  Our results,
averaged over all clusters but segregated by spectral type, are shown
in Fig.~\ref{fig:dsave}. The average DS values at different $N_{obj}$
for each type are shown as offsets from the DS values averaged over
all galaxy types.

This analysis indicates that the LCBGs show a stronger signal for
subclustering at the smallest scales as compared to the other galaxy
types.  We interpret this as indicating that LCBGs are preferentially
found either in an interacting pair or a small group.  Likewise, the
blue galaxies, of which the LCBGs comprise a significant number, show
a positive signal for subclustering at small scales.  In
contrast, the red galaxies show no sign of subclustering at small
scales and the GV galaxies are actually strongly anti-clustered at the
smallest scales ($N_{obj}=5$) compared to the overall cluster
population.  At larger scales ($N_{obj}>12$), the red and GV galaxies
show stronger indications of being a part of cluster substructure
(i.e., the clusters themselves), whereas the blue galaxies exhibit a
more uniform distribution.

\begin{figure}
\epsscale{1.0}
   \includegraphics[scale=0.7,angle=270]{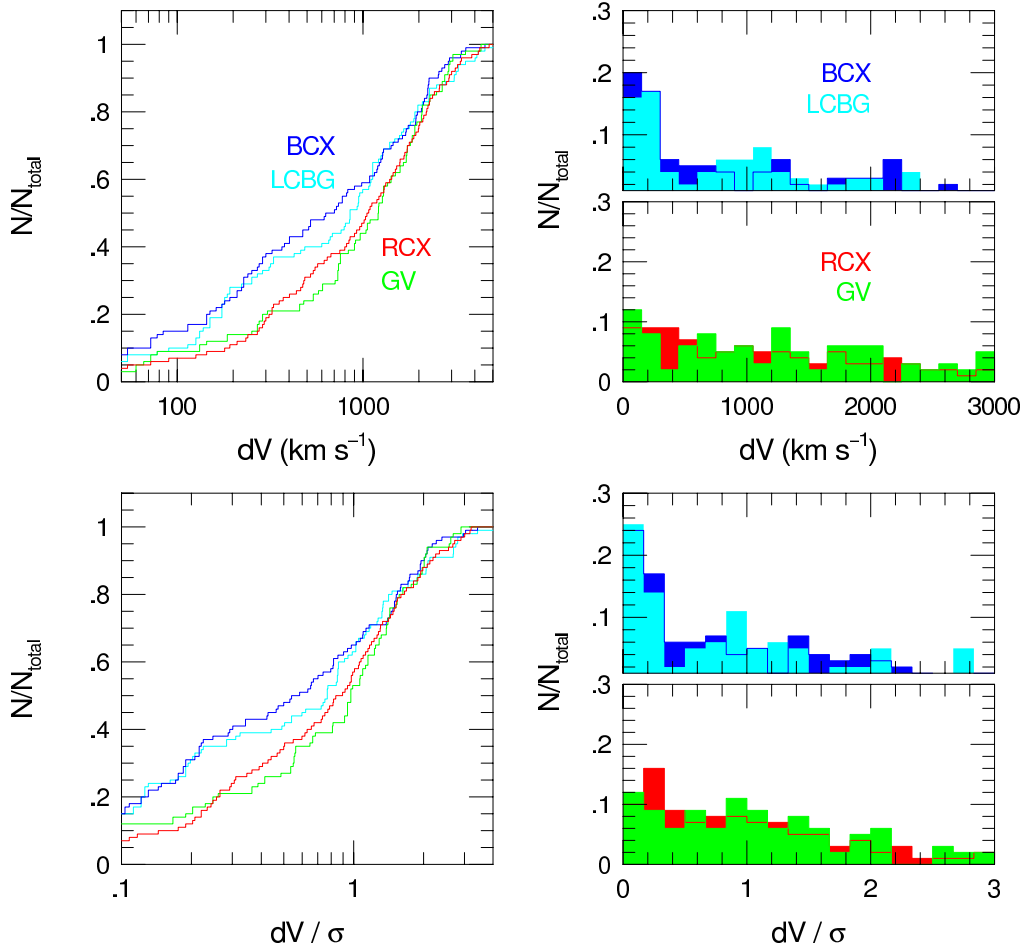}
  \caption{
    \label{fig:neighbor}
    Cumulative (left) and differential (right) histograms of velocity
    differences (dV) for bonafide cluster members with respect to
    their spatially-projected nearest neighbor. Histograms combine
    data for all clusters with no weighting, are broken down by
    spectral type, and are normalized by the total number of galaxies
    in all clusters for all spectral types.  Bottom panels show the
    same histograms with the velocity difference normalized by the
    global cluster velocity dispersion (dV/$\sigma$) relevant for each
    galaxy. Color coding by spectral type is the same as in previous
    figures (see labels in top panels).}
\end{figure}

To further explore the small scale clustering of LCBGs, we considered
the relationship between a given galaxy's radial velocity and that of
its nearest neighbor (viewed in projection on the sky).
Figure~\ref{fig:neighbor} depicts the normalized distribution of the
velocity difference between each galaxy and its nearest neighbor.  In
this plot, many more galaxies in the LCBG and BCX classes show small
velocity differences with their closest neighbor, indicating that
these galaxies are more likely to have radial velocities correlated
with those of their nearest neighbor.  The K-S test rejects the
hypothesis that the distribution in nearest-neighbor velocities for
LCBGs is drawn from the same distribution as RS galaxies.  The
pairwise velocity differences for the GV and RSX galaxies are
comparable to the cluster velocity dispersion, while the same quantity
for the LCBG and BCX types is substantially lower.  In particular, the
LCBG and BCX samples have a significant enhancement of galaxies at low
dV, namely at $< 30\%$ of the cluster or roughly 300~\kms , comparable
to the characteristic binding energy of a typical galaxy.  This
enhancement is a factor of two relative to GV and RSX, and $40\%$ of
the total fraction for each LCBG and BCX subpopulation.

The results of our clustering analysis might depend on
where we draw the photometric boundaries between different populations
because these boundaries are intended to delineate galaxies with
distinct stellar populations. However, the differences should not be
apparent when boundary shifts are made on scales comparable to the
photometric errors (0.05 mag or less). We can test this with the green
value (GV) sample that straddles the region in color-magnitude space
(here rest-frame $U-B$ and $M_B$) between the red sequence and the
blue cloud. Specifically, GV galaxies are defined to lie in a 0.2 mag
wide band in $U-B$ that changes linearly with absolute magnitude
(\S3). Since the GV clustering statistics are similar to the RS
clustering statistics, we expect that shifting the color-band to the
red will not change the cluster statistics. Shifting the color-band
sufficiently to the blue (0.2 mag) should change the GV clustering
statistics to more closely resemble those of the BC. The rapidity of
this change will depend on the uniformity of the clustering statistics
within the BC sample, which subtends a broader color range than the GV
sample. The relative similarity between BC and LCBG clustering
statistics suggests the clusering statistics within the BC sample are
rather uniform.

Accordingly, we've repeated the substructure measurements for the GV
sample with 5\%, 10\% and 20\% changes to their selection criteria,
i.e., 0.05, 0.10 and 0.2 mag shifts to the red and the blue of the
color-band in $U-B$ that defines the GV class.  All of the shifts of
the color-band to the red yield clustering statistics consistent with
that of the nominal definition, as we expect. For shifts of the
color-band to the blue, shifts of 0.05 mag yield no change; shifts of
0.10 mag yield clustering statistics intermediate between GV and BC
samples; while 0.2 mag shifts yield clustering statistics comparable
to the BC sample. We conclude, then, that the transition in the
clustering properties between BC and RS/GV samples is rather rapid in
color (0.10--0.15~mag), and robust to photometric errors.

\section{Discussion}

\label{discussion} In this work, we have presented the spatial and
kinematic distributions of several different galaxy populations in
massive, intermediate-redshift clusters.  As expected, the majority of
cluster galaxies belong to the red sequence (RS) class.  These have a
compact distribution and show velocity and spatial distributions
similar to low-redshift, massive clusters.  The clusters also have a
large fraction of blue (BC) galaxies as compared to lower redshift
clusters. This blue population tends to be more extended and less 
centered than any of the other populations, but has a
significant amount of subclustering, and velocity
distribution boxier than the RS.

We focus our attention here on the two transitional classes of
galaxies, LCBGs and GV galaxies, which have spatial extents
intermediate between the BC and RS populations, but velocity
distributions that are boxier than the BC population.  The LCBGs
and GV galaxies have different subclustering amplitudes comparable to
the BC and RS populations, respectively.  In addition, the LCBGs are
distinguished by a slightly enhanced velocity dispersion.

\subsection{The Distribution of LCBGs}

The number density of LCBGs is enhanced in clusters, indicating that
the cluster environment plays some role in triggering the starburst
\citep{cra06,cra11}; this effect has also been observed for
$24~\micron$ sources around clusters at intermediate redshift
\citep{marcillac07, fadda08, geach11, noble13} and for
optically-identified star-forming galaxies in low- and
intermediate-redshift clusters \citep{balogh04, moss06, porter07,
reverte07, porter08, oemler09, mahajan12}.  A key aspect in
understanding the evolution of cluster LCBGs and star-bursting
galaxies is determining what triggers this phase of star formation.

We have presented several pieces of evidence to indicate that the
LCBGs represent a distinct cluster population:
\begin{enumerate}

\item LCBGs exhibit a radial velocity distribution which differs from
that of RS galaxies, showing greater kurtosis.

\item LCBGs in these clusters have, on average, a higher velocity
dispersion than the other populations.

\item The spatial distribution of LCBGs differs from the RS galaxies,
since the latter are found in the cluster core while the former are
not.

\item LCBGs are more likely to occur in pairs or small groups as
indicated by our subclustering analysis.

\item LCBGs are more likely to have a velocity similar to their
nearest neighbors in projected distribution on the sky.

\end{enumerate}
The differences in velocity and spatial distribution indicate a
population of galaxies that has not yet achieved equilibrium within
the cluster.  This distribution suggests that the LCBG class
constitutes an infalling population and does not show the
centrally-peaked velocity distribution expected for a backsplash
population \cite[\ie, a population that has passed through the central
region of the cluster at least once;][]{gill05}.

For an infalling population, several possible triggers for star
formation have been postulated: interaction with the ICM
\citep{fujita99}, tidal forces within the cluster \citep{bekki99},
mergers of large subsubstructure \citep{bekki10}, and tidal
interactions with other galaxies \citep{gnedin03}.  In all likelihood,
starbursts in clusters result from a diversity of causes, with each
of the different methods contributing to the burst seen in the galaxy.
We can examine the likelihood and evidence for the contribution of
each of these different methods to the triggering of the LCBG phase.

\cite{fujita99} suggested, based on hydrodynamic simulations, that
star formation initially increases as a galaxy approaches the cluster
due to compression from the ICM and then drops due to stripping from
the ICM within 1 Mpc.  Likewise, \cite{kronberger08} used simulations
to find that ram pressure stripping can also increase star formation
rates in galaxies by up to a factor of 3 and also create ``stripped
baryonic dwarf galaxies.''  The pressure at which these processes
becomes effective is much less than that required to strip the galaxy
\citep{bekki03}, and so they can affect the star formation out to much
larger radii.  In \wx, we find evidence for ``walls'' of LCBGs which
may be associated with shock fronts in the cluster ICM as proposed
by \cite{bekki10}.  However, further investigation of the ICM
properties in this cluster are necessary to confirm these findings.

Clusters are expected to gain up to 40\% of their stellar mass via
mergers with groups that have masses greater than $10^{13} h^{-1}
\mbox{M}_{\odot}$ \citep{mcgee09}.  Simulations indicate that
starbursts could be induced via tidal forces during the merger process
in member galaxies of the group \citep{bekki99} as well as
interactions with shocks in the ICM causing synchronized bursts of
star formation \citep{bekki10}.  However, in only one of our clusters
(\wov) do we find evidence that LCBGs are associated with large scale
($N_{obj}>10$) substructure.  On the other hand, within that
substructure we do observe a very large fraction of LCBGs.  In
simulations, \cite{cohn12} showed that only half of the simulated
clusters experienced mergers with large groups, although those
subgroups did tend to survive for long periods of time (4~Gyr for
50\% disruption, albeit the DS test was unable to reproduce the
substructure the majority of the time).  From our small sample of
clusters, it is difficult to establish a correlation between LCBGs and
substructure.
                    
\cite{gnedin03} has shown that tidal heating, especially from close
galaxy encounters, can substantially contribute to the heating of
galaxies in a cluster.  Furthermore, \cite{tonnesen12} found in their
simulations that bound pairs in high density regions had higher
fractions of star-forming galaxies and that pair-bound galaxies also had a
higher specific star-formation rate than the galaxy population as a
whole.  The pronounced subclustering of LCBGs at the smallest scales
could be evidence that this is the dominant phase for triggering the
starbursts.  Of 89 LCBGs in our sample, 16 LCBGs have a confirmed
nearby neighbor (within 100~kpc and 150~\kms) with an additional 37
having a potential nearby neighbor (any source within 25~kpc).  In
their sample of low redshift LCBGs, \cite{perez11} find 45\% of their
sample to have a nearby companion within 25~kpc, which is similar to
our fraction of potential neighbors.

\subsection{Green Valley Galaxies}

Our observations show that the phase-space distribution of GV galaxies
is both similar to and different from the phase-space distribution of
RS galaxies.  GV galaxies have a similar mean velocity and velocity
dispersion as the RS galaxies, but the distribution is boxier (\ie,
having lower kurtosis) than the RS distribution.  The subclustering of
the two populations has the same behavior over different scales, yet,
unlike the RS galaxies, GV galaxies appear to be completely absent
from the cores of the clusters.

If we consider a simple model in which GV galaxies evolve from
quenched BC galaxies that fall in along radial orbits, this constrains
the timescale for the GV phase to a maximum of $\sim1$~Gyr
\citep{bg06}.  This also corresponds to the typical time for galaxies
to move from the blue cloud to the red sequence \citep{martin07}.
Thus, the quenching of the galaxies has to begin near or beyond the
virial radius, which would closely correspond to the onset of
ram-pressure stripping in these clusters \citep{moran07}.  The GV
phase transition, expected to last only 100--200~Myr \citep{martin07},
would occur prior to the galaxy reaching the central cluster core.

However, this model is probably overly simplistic.  In our massive
clusters, the ratio of GV to BC galaxies is 0.3, which in this simple
model would indicate either that the GV phase lasts closer to 300~Myr
or that the infall rate was higher in the past.  Other than the cored
profile, the GV galaxies share many similarities in the spatial and
velocity distribution as the RS galaxies, particularly in terms of
subclustering. This indicates that the GV galaxies have been in the
clusters for a longer time period than the BC population and have made
multiple cluster crossings (assuming radial orbits). \cite{wetzel13}
find that satellite galaxies are in clusters for 2-4 Gyrs prior to a
rapid quenching period, which would be consistent with our results
here.

Furthermore, the GV galaxies do not necessarily show signs of being a
backsplash population \citep{gill05}, but our observations do not
reach beyond the virial radius where differences in line of sight
velocity may further distinguish between virial, infall, and
backsplash populations \citep{mahajan11}. The GV do show an
extended spatial distribution with many appearing on the outer edges
of the cluster, suggesting that the quenching process may have started
much earlier or been far more effective at transforming BC galaxies to
GV galaxies.  A picture consistent with our observations is one where
GV galaxies have long been cluster members, but are preferentially on
larger and more circular orbits which enable them to retain and
convert more of their gas into stars over a longer time than their RS
counterparts.

In the field at low redshift, GV galaxies constitute approximately
10\% of the population \citep{chen10}.  Of these, up to $75\%$ of the
GV population show signs of AGN activity \citep{martin07}, with many
of those only being X-ray detected AGN \citep{hickox09}.  Further
multi-wavelength studies will be required to assess whether or not
cluster GV populations share a similar quenching mechanism with the
field population.
     
\section{Conclusions}
\label{sxn-summary}

We have explored the projected phase-space distributions of different
classes of spectroscopically-confirmed members in five massive,
intermediate-redshift clusters. The subclasses are defined by
multi-band photometry and correspond to reproducible definitions
comparable to those found in the literature for red sequence (RS),
blue cloud (BC), green valley (GV), and luminous blue compact galaxies
(LCBG). The GV class straddles the locus separating the BC and RS
subclasses, and the LCBG class is a subset of the BC sample with
high surface brightness and small size.  We have further defined BCX
and RSX subclasses identical to BC and RS, respectively, except for
the exclusion of GV and LCBG galaxies.

We have employed a technique (a modification of the ``shifting
gapper'') which allows us to identify cluster members objectively and
with minimum ambiguity.  In accordance with previous studies, we find
these distant clusters to harbor a larger fraction of BC galaxies
compared to lower-redshift counterparts of comparable mass ($>10^{15}
\ M_{\odot}$). In these massive, compact clusters, the fraction of
galaxies in each subclass corrected for incompleteness down to a fixed
$B$-band absolute magnitude of $M_B=-18.5$~mag over the redshift range
of our clusters is: $61\pm3$\% (RSX), $11\pm4$\% (GV), $26\pm6$\%
(BCX), and $19\pm3$\% (LCBG).

Based on these spectroscopically-confirmed cluster members, we have
measured the projected spatial and velocity distribution of all galaxy
types in each cluster, including updated measurements of the velocity
dispersion for each of the clusters.  We have also measured the
clustering properties of each galaxy subclass (based on \cite{ds88})
over a wide range of clustering scales (following \cite{cd96}). The
clustering properties and projected phase-space distribution of each
galaxy subclass relative to the cluster barycenters show distinct
patterns. This correlation of photometric, spatial, kinematic, and
clustering properties hints at a causal connection between the
star-formation history of cluster populations, their infall, and
possibly their orbital structure.

Our spectroscopic observations and cluster-membership analysis confirm
previous findings based solely on photometric data \citep{cra06} that,
in contrast to the RS galaxies, BC and LCBG populations avoid the
cluster core. The analysis here extends the finding of core-avoidance
to the GV population. With these facts alone, it is tempting to link
the LCBG and GV populations to subsets of the BC population that are
in different phases of structural and dynamical transition from an
infalling, starbursting population to an equilibrium, quiescent RS
population.  Specifically, we find all three classes (BCX, LCBG, and GV) have
greater velocity kurtosis, spatial offset, and radial extent than RSX,
with GV and LCBG being matchingly extreme in velocity kurtosis and
spatial ellipticity.  However, this simple interpretation is
contradicted by clustering analysis which clearly links the LCBG
galaxies to the BCX subclass by virtue of their close association with
their nearest neighbor and small-scale structure. In distinction, the
GV population has subclustering properties similar to the RSX
population dominating the cluster cores.

These findings lead us to conclude the following 
regarding the transitional populations in these clusters:

\begin{itemize}
  
\item GV galaxies are long-lived cluster members that may be on
  preferentially circular and larger-radius orbits than their RSX
  counterparts. Such orbits would help GV galaxies retain more of
  their gas and perpetuate star-formation relative to RSX galaxies.
  While this interpretation is not unique, added support comes from
  the fact that a simple infall model is not sufficient to fully
  explain the properties of the GV galaxies.
  
\item LCBGs are extreme in their blue color, high surface-brightness
  and small-scale clustering amplitude even within the BC population.
  These characteristics, combined with their systematically larger
  velocity dispersion compared to other cluster subclasses, depict
  LCBGs as a radially-infalling population with star formation
  enhanced by group interaction and impact with the intracluster
  medium.

\end{itemize}
In future work, we will probe the spectroscopic properties of these
objects with the goal of obtaining definitive answers about the
star-formation timescales and dynamical masses of the transition
populations within these massive clusters.

\acknowledgments

%% To help institutions obtain information on the effectiveness of their
%% telescopes, the AAS Journals has created a group of keywords for telescope
%% facilities. A common set of keywords will make these types of searches
%% significantly easier and more accurate. In addition, they will also be
%% useful in linking papers together which utilize the same telescopes
%% within the framework of the National Virtual Observatory.
%% See the AASTeX Web site at http://www.journals.uchicago.edu/AAS/AASTeX
%% for information on obtaining the facility keywords.

We thank the referee for the careful reading of our manuscript
and the constructive criticism that improved our paper.  
S.M.C. acknowledges the South African Astronomical Observatory and the
National Research Foundation of South Africa for support during this
project.  M.A.B. acknowledges suppport from NSF grant
AST-1009471.  This work made use of IRAF, a software package
distributed by the National Optical Astronomy Observatory, which is
operated by the Association of Universities for Research in Astronomy
(AURA) under cooperative agreement with the National Science
Foundation.  The authors wish to recognize and acknowledge the very
significant cultural role and reverence that the summit of Mauna Kea
has always had within the indigenous Hawaiian community.  We are most
fortunate to have the opportunity to conduct observations from this
mountain.

%% After the acknowledgments section, use the following syntax and the
%% \facility{} macro to list the keywords of facilities used in the research
%% for the paper.  Each keyword will be checked against the master list during
%% copy editing.  Individual instruments or configurations can be provided 
%% in parentheses, after the keyword, but they will not be verified.

{\it Facilities:} \facility{WMKO, WIYN}

\end{document}